\documentclass{emulateapj}
\usepackage{graphicx}
\usepackage{natbib}
\usepackage{multirow}

\def\apjref#1;#2;#3;#4 {\par\pni\ #1,  #2, {\bf #3}, #4. \par}

\newcommand{\beq}	{\begin{equation}}
\newcommand{\eeq}	{\end{equation}}
\newcommand{\beqa}{\begin{eqnarray}}
\newcommand{\eeqa}{\end{eqnarray}}
\newcommand{\avg}[1]  {{\langle #1 \rangle}} 

\newcommand{\e}	{$^{-1}$}

\def\simlt{\lower.5ex\hbox{$\; \buildrel < \over \sim \;$}}
\def\simgt{\lower.5ex\hbox{$\; \buildrel > \over \sim \;$}}
\def\la{\simlt}

\font\tenbi=cmmib10 
\newfam\bifam  \textfont\bifam=\tenbi

\font\tenbr=cmbx10
\newfam\brfam  \textfont\brfam=\tenbr

\font\squinttenbi=cmbx10 at 9pt
\scriptfont\brfam=\squinttenbi

\def\vecnabla{
             \setbox1=\hbox{$\bigtriangledown$}
                          \raise.45ex\hbox{$\bigtriangledown$\hskip-.97\wd1
                          $\bigtriangledown$\hskip-.97\wd1
                          $\bigtriangledown$\hskip-.97\wd1}
                          \raise.47ex\hbox{$\bigtriangledown$}}

\newcommand{\caln}		{{\cal N}}
\newcommand{\calr}		{{\cal R}}

\newcommand{\ppbyp}[2]	{{{\partial#1}\over{\partial#2}}}
\def\rsun{\ifmmode {R}_{\mathord\odot}\else {\it R}$_{\mathord\odot}$\fi}
\def\msun{\ifmmode {M}_{\mathord\odot}\else {\it M}$_{\mathord\odot}$\fi}
\def\lsun{\ifmmode {L}_{\mathord\odot}\else {\it L}$_{\mathord\odot}$\fi}

\newcommand{\jf}		{{j_f}}

\newcommand{\Lsun}	{L_\odot}  
\newcommand{\mf}		{{m_f}}
\newcommand{\ml}		{{m_\ell}}
\newcommand{\mfl}		{{m_{f,\ell}}}

\newcommand{\mmax}	{{m_{\rm max}}}
\newcommand{\mup}		{{m_u}}

\newcommand{\mdone}		{\dot m_1}
\newcommand{\mdo}		{\dot m_0}
\newcommand{\mdoo}	{\dot m_{0,\,1}}
\newcommand{\mdot}	{\dot m}
\newcommand{\mds}		{\dot m_{\rm IS}}
\newcommand{\mdtc}	{\dot m_{\rm TC}}
\newcommand{\mdca}	{\dot m_{\rm CA}}
\newcommand{\nds}		{\dot\caln_*}

\newcommand{\ppt}		{\psi_{p2}}

\newcommand{\rmd}		{\calr_{{\dot m}}}
\newcommand{\rmdca}		{\calr_{{\dot m,\,\rm CA}}} 
\newcommand{\scl}		{\Sigma_{\rm cl}}

\newcommand{\tfo}		{t_{f1}}

\newcommand{\tf}		{t_f}
\newcommand{\tfobs}		{\avg{t_f}_{\rm obs}} 

\newcommand{\facc}		{f_{\rm acc}}
\newcommand{\lacc}		{L_{\rm acc}}
\newcommand{\lmin}		{L_{\rm min}}

\newcommand{\mlmf}		{m(L,\mf)}
\newcommand{\mflm}		{\mf(L,m)}

\newcommand{\rmmd}		{r(m,\dot m)}
\newcommand{\Ppl}		{\Psi_p(L)}
\newcommand{\Ppt}		{\Psi_{p2}}

\newcommand{\mmin}	{{m_{\rm min}}}

\newcommand{\mfellm}            {\mf(\ell,m)}
\newcommand{\mdis}              {\dot m_{\rm IS}}

\shorttitle{The Protostellar Luminosity Function}
\shortauthors{Offner \& McKee}

\begin{document}

\title{The Protostellar Luminosity Function}

\author{Stella S. R. Offner}
\affil{Harvard-Smithsonian Center for Astrophysics,
Cambridge, MA 02138 }
\email{soffner@cfa.harvard.edu }
\and
\author{Christopher F. McKee}
\affil{Physics Department and Astronomy Department, University of California,
  Berkeley, CA 94720}
\email{cmckee@astro.berkeley.edu}

\begin{abstract}
The protostellar luminosity function (PLF) is the present-day luminosity
function of the protostars in a region of star formation. It is
determined using the protostellar mass function (PMF) in combination
with a stellar evolutionary model that provides the luminosity as a
function of instantaneous and final stellar mass. 
As in McKee \& Offner (2010), we consider three
main accretion models: the Isothermal Sphere model, 
the Turbulent Core model, 
and an approximation of the Competitive Accretion model. 
We also consider
the effect of an accretion rate that tapers off linearly in time and
an accelerating star formation rate. 
For each model, we characterize
the luminosity distribution using the mean, median,
maximum, ratio of the median to the mean, standard deviation of the
logarithm of the luminosity, and the fraction of very low luminosity
objects. We compare the models with bolometric luminosities observed in local
star forming regions and find that 
models with an approximately
constant accretion time, such as the Turbulent Core and Competitive Accretion models,
appear to agree better with observation than those with a 
constant accretion rate, such as the Isothermal Sphere model.
We show that observations of the mean protostellar luminosity
in these nearby regions 
of low-mass star formation 
suggest a mean star formation time of 0.3$\pm$0.1 Myr.
Such a timescale, together with 
some accretion that occurs
non-radiatively 
and some that occurs
in high-accretion, episodic bursts, resolves the classical ``luminosity problem"
in low-mass star formation, in which observed protostellar luminosities are
significantly less than predicted.
An accelerating star formation rate is one possible way of reconciling
the observed star formation time and  
mean luminosity.
Future observations will place tighter constraints on
the observed luminosities, star formation time, and episodic
accretion, enabling better discrimination between star formation
models and clarifying the influence of variable accretion on the PLF.
\end{abstract}

\section{Introduction}

Protostars are born in dense, compact molecular cloud cores \citep{mckee07}. During their earliest evolution, protostars are both dim and heavily obscured by a dusty
envelope. It is an unavoidable observational reality that the majority of the star formation process
occurs while protostars are deeply embedded within their natal gas.
Consequently, high extinction and significant radiation
reprocessing render the details of protostellar evolution, lifetimes, and the
accretion process extremely uncertain \citep{white07}.

Recent surveys of the nearby star-forming regions have successfully
obtained large statistical samples of young protostars with reasonable
completeness down to
luminosities of $\sim 0.1~\lsun$ (e.g., \citealt{evans09, enoch09, dunham08}). High-resolution millimeter emission maps tracing dusty envelopes,
provide a measure of core masses \citep{enoch08}. Combined with mid to
far-infrared
data, the available wavelength range is sufficient to trace the spectral energy
distribution (SED), from which the total luminosity may be estimated.

Using the infrared spectral slope, observed sources can be divided into four classes that can be approximately mapped to evolutionary stages
\citep{andre00}. 
Class 0 protostars are heavily obscured by a dusty
envelope, 
such that most of the radiation falls in the sub-millimeter band. During the
Class I phase the protostar, 
while still embedded, becomes less obscured and may be surrounded by a 
thick circumstellar accretion disk.  
By the Class II phase, the now pre-main sequence star has accreted or 
expelled most of the initial envelope mass and produces little sub-millimeter emission. 
The remaining gas lies in thin accretion disk surrounding the star. 
Signatures of outflows may be apparent during both the Class I and
Class II phases.  
During Class III, the disk dissipates and the source approaches the
main sequence.

Despite this straightforward picture, cataloging sources and definitively mapping them to a
physical stage remains 
complicated. Geometric effects shift objects over the Class 0/I
boundary, distorting the correlation between physical stage and SED
characteristics.  
Edge-on Class II protostars with higher extinction may be misclassified as
Class I, while Class I sources viewed down the outflow cavity resemble
Class II sources \citep{masunaga00, robit06}. 
Variability in the accretion rate may cause the protostar to oscillate
across class boundaries \citep{dunham10}.
Measurements of the millimeter emission used as a proxy for the
envelope mass can be used to distinguish between embedded, i.e. Class 0 and Class I
objects, 
and non-embedded, Class II objects \citep{enoch08}, although objects
with thick disks along the line-of-sight may still be misclassified.

During the Class 0 and I phases,
luminosity due to accretion likely dominates the total
radiative output \citep{evans09,dunham10}. Consequently, upper limits
for the accretion rates may be inferred from the luminosity \citep{enoch09}. 
This information gives clues about the formation timescale and the accretion
process 
during which the protostars are deeply embedded and cannot be directly
imaged. However, the large distribution of observed luminosities, and in particular
the significant number of dim protostars, creates a picture that is
largely inconsistent with predictions of some star formation models.

\subsection{The Luminosity Problem}

The accretion rates inferred and the star formation
times 
predicted by theoretical models combine to suggest that
protostars are on average too dim (e.g.,
\citealt{kenyon90,young05,enoch09}). 
The typical accretion luminosity
of a protostar
\beqa
\lacc&=&\facc\left(\frac{Gm\dot m}{r}\right),
\label{eq:lacc}\\
&=& 7.8\facc\left(\frac{m}{0.25 M_\odot}\right) \;\times
\nonumber\\
&&~~~~~\left(\frac{\dot m}{2.5\times 10^{-6} M_\odot\;\mbox{yr\e}}\right)\left(\frac{2.5~\rsun}{r}\right)~\lsun
\label{eq:lacc2}
\eeqa
exceeds the observed average luminosity of about $2~\lsun$ and significantly
exceeds the observed median luminosity of $0.9~\lsun$ \citep{enoch09}. Here
we have adopted a typical mass accretion rate of $\dot m=2.5\times 10^{-6}M_\odot$~yr\e\
\citep{mckee07} and a typical protostellar radius
of $r=2.5~\rsun$ \citep{stahler88,hartmann97}. 
The factor $\facc$ is the fraction of the gravitational
potential energy that is radiated away. Note that $\lacc$ is the 
total accretion luminosity, including
any emission from an accretion disk around the star.
There are three factors that can reduce
the accretion efficiency $\facc$ below unity: First, some of the energy 
of  the gas that accretes onto the protostar can be extracted by
a hydromagnetic wind (e.g., \citealp{ostriker95}).
If half the energy lost by the disk is mechanical instead of radiative, then,
since half the total potential energy is lost by the disk, this would give 
$\facc\simeq \frac 34$. Second,
some of the energy in the in-falling gas can be absorbed by the star;
\citet{hartmann97} argue that this is not significant for accretion rates $\la 10^{-5} M_\odot$~yr\e,
and this has been confirmed by \citet{commercon11}
so long as the accretion flow is transparent to optical radiation.
Finally, some of the accretion energy is used in dissociating and then ionizing the
accreting gas \citep{tan04}, but this is significant only for very low masses.
Thus, of these three effects, the mechanical loss of energy is the only one of significance. 
As we shall see in Section \ref{sec:epi}, episodic accretion can reduce the observed
mean luminosity below the true mean and therefore contribute an effective reduction
in $\facc$.

\subsection{Review of Past Work}
A number of mechanisms have been
suggested to resolve the luminosity problem. A successful solution must account for both
the mean luminosity of the distribution and the spread of luminosities over several orders of magnitude.
The first indication of the discrepancy between accretion and
luminosity was observed in Taurus by \citet{kenyon90, kenyon95}. At that time, observations of
only a handful of young protostars existed, and the authors speculated
that the problem could be resolved 
by, among others,
short, unobserved periods of
high accretion or increased ages of T Tauri
stars.

\citet{fletcher94a,fletcher94b} performed the first derivations of
protostellar mass and luminosity functions. Assuming constant accretion rates
and a fixed time period during which stars formed with a constant star formation rate,
they derived the time-dependent luminosity function for a cluster of evolving
protostars. Their model included the luminosity contribution from
Class I to main-sequence stars.
For steady-state star formation, our constant accretion-rate
model produces a protostellar protostellar luminosity function similar
to that derived by \citep{fletcher94b} when pre-main sequence stars
are excluded (see \S\ref{s.plf}).

\citet{myers98} used a semi-analytic model to describe the evolution of the protostar, disk, and envelope system. With simple radiative transfer, they generated
bolometric temperature and luminosity tracks from Class 0 to the zero-age main sequence stage. 
In order to
represent the decline in accretion luminosities with time, the models assumed an exponentially
decreasing accretion rate.
For some parameters, the $L-T$ tracks were able to reproduce the mean
protostellar luminosities. 
However, the breadth of the distribution and the relative number of low-luminosity sources were not addressed.

Monte-Carlo radiation transfer modeling by \citet{whitney03} underscored the
the dependence of the SED on core geometry. In particular, they showed that the
inferred bolometric luminosity may vary by 50\% from the true
luminosity depending upon the orientation of the disk and outflow cavity
relative to the line-of-sight. 
For a sufficient number of observed protostars
sampling all geometric projections, the orientation can only alter the distribution of the
luminosities, not the mean. 

\citet{young05} performed 1-D
radiative transfer observations of the inside-out collapse model developed by \citet{shu77}. These calculations improved upon the \citet{myers98} work by following the hydrodynamic evolution of the protostellar core.
They included six contributions to the total luminosity with the result that their $L-T$ tracks agreed
only with the brightest protostars, further illuminating the discrepancy
between observed luminosities and predictions of low-mass star
formation.

Recent extensive surveys of five local star-forming regions have provided more comprehensive statistics and observational data of the earliest stages of protostellar evolution \citep{enoch09, evans09}
This work demonstrated that Class 0 and Class I sources have
comparable mean luminosities, suggesting that protostars may accrete
significantly for longer periods of time than previously
assumed. Extending the accretion epoch from 0.1 Myr, as first assumed
by \citet{kenyon90}, to 0.5 Myr presents one possible solution to the luminosity problem.

\citet{dunham10} revisited the \citet{young05} methods and explored
several improvements, including updated opacities, 2-D effects such as
disks, core rotation,  outflow cavities, and variable
accretion. 
With all the additions, although predominantly due to the inclusion of
episodic accretion, good
statistical agreement between the observed and modeled temperature and luminosity distributions
was possible.
They modeled variability by halting accretion from the
disk onto the protostar until the disk mass reached 20\% of the star
mass, whereupon the star was allowed to accrete at a constant rate of $10^{-4}
~\msun$ yr$^{-1}$ until the disk mass was exhausted.  Since other terms
contributing to the total luminosity (e.g., accretion from the
envelope onto the disk and the stellar luminosity) are small compared
to the accretion luminosity, the model assumes that protostars exist in a very low
luminosity state for most of their formation time and accrete most of
their mass
during bursts.
Statistically, their models predict that protostars must spend $\sim 1$\%
of the time radiating at $L>100\lsun$. This is only marginally consistent with the
observed sample of 112 protostars, which contains no sources more luminous than
$76~\lsun$. The models also have a star formation time $<0.2$Myr, which
is less than the star formation time inferred by \citet{evans09}.
Future surveys containing more objects are necessary
to test the importance of episodic accretion.

There is debate whether
allowing for episodic accretion in evolutionary models can reproduce the dispersion of
low-luminosity sources on the Hertzsprung-Russell diagram and thus
account for a portion of the age spread inferred for members in young, low-mass star
clusters  \citep{baraffe09, hosokawa11}.  If so, episodic
accretion could also explain lithium depletion measurements in young stars
\citep{baraffe10}.
However, this solution is valid only if
nearly all the accretion energy is efficiently radiated away from a
small fraction of the
stellar surface. 
Other
mechanisms such as varying initial conditions and assumptions about
the radiative properties of the accretion flow, which are not well
constrained, may also contribute to an apparent age spread \citep{hosokawa11}.

\subsection{Model Assumptions}

In this paper we derive the Protostellar Luminosity Function (PLF)
predicted by various star formation models and compare 
with observations \citep{enoch09}. Comparison
with the mean observed luminosity determines the mean star formation time;
comparison with the shape of the PLF (ratio of median to mean and the 
standard deviation) tests the validity of the models. Our basic
formalism (McKee \& Offner 2010a, henceforth Paper I)
can be adapted to consider 
a time-varying rate of star formation, bursty accretion,and an
arbitrary stellar initial mass function. However, in this paper we
base our comparison on  
four
main assumptions.
First, we treat only the cases in which the star formation rate is either constant or smoothly
accelerating in time (e.g. \citealt{palla00}). Second, we assume the accretion rate is a smooth function of time during the accretion phase;
in particular, we assume it can be expressed as a function of the current mass and the final
mass of the protostar. Rather than excluding stochastic variability
completely, we assume that any time-varying component is
statistically rare in the samples with which we compare and therefore
unlikely to be included the data (see \S5.3 for a detailed discussion).
Third, we assume that the accretion rate onto the protostar, which can be inferred
from the protostellar luminosity, tracks the accretion rate onto the protostar-disk
system over the majority of the lifetime of the protostar.
Finally, we adopt 
an individual-star
\citet{chabrier05} IMF truncated at an upper mass limit $\mup$.

Through our derived PLF models, we investigate the
luminosity problem in the context of different star formation
theories and explore variations such as an accelerating star 
formation rate and accretion rates that taper off over time.  
In \S\ref{s.pmf} we review the Protostellar Mass Function, which is
described in detail in Paper I. 
In \S\ref{s.plf} we derive
the PLF and relative statistics. We compare with
observations of local star-forming regions in \S\ref{s.comp}
and discuss our results in \S\ref{s.dis}.
Approximate analytic results for the PLF for different star formation models
are given in the Appendix.

\section{The Protostellar Mass Function (PMF)} \label{s.pmf}

In Paper I we determined the mass function of protostars (the PMF) in
terms of the IMF and the accretion history of the protostars.
We assumed that the accretion history could be expressed in terms of
the current protostellar mass, $m$, and its final, stellar mass,
$\mf$. Henceforth, $m$ and $\mf$ are measured in units of solar masses
and $\mdot$ is in units of solar masses per year.
We expressed the IMF as $\psi(\mf)$, where $\psi(\mf)d\ln\mf$ is
the fraction of stars born with final masses in the range $d\mf$. We then
defined the bi-variate PMF, $\ppt$, such that $\ppt\, d\ln m\,d\ln\mf$ is
the fraction of protostars in a region of star formation with
masses in the range $dm$ and final masses in the range $d\mf$.
We showed that for steady (i.e., non-accelerating) star formation
\beq
\ppt(m,\mf)=\frac{m\psi(\mf)}{\dot m\avg{t_f}},
\label{eq:ppt}
\eeq
where $t_f$ is the formation time for a star of mass $\mf$ and
$\avg{t_f}$ is the average of $t_f$ over the IMF.
The PMF, $\psi_p(m)$, is just the integral of the bi-variate PMF,
\beq
\psi_p(m)=\int_\mfl^\mup \ppt(m,\mf) d\ln\mf,
\label{eq:pmf}
\eeq
where the lower limit of integration is
\beq
\mfl=\max(\ml,m)
\label{eq:mfl}
\eeq
and the most and least massive stars formed by the cluster are
$\mup$ and $\ml$, respectively.

To determine the PMF, we required the accretion histories of the protostars.
As noted above, we assume that these accretion histories, apply to
the gas reaching the protostellar surface as well as to the gas that
accretes from the ambient medium onto the protostar-disk system.
We considered several different models:
\begin{itemize}
\item[i)] Inside-out collapse of an isothermal sphere (IS, \citealt{shu77}),
\beq
\dot m= \mds=1.54\times 10^{-6} (T/10\;\mbox{K})^{3/2}.
\label{eq:mds}
\eeq
\item[ii)] The Turbulent Core model (TC, \citealt{mckee02,mckee03}), in which the
initial core is presumed to be supported by turbulent motions instead of
thermal pressure,
\beqa
\dot m&=&3.6\times 10^{-5}\scl^{3/4}\left(\frac{m}{\mf}\right)^j \mf^{3/4},\\
\label{eq:mdtc}
&\equiv& \mdtc \left(\frac{m}{\mf}\right)^j \mf^{3/4},
\eeqa
where $\dot m$ is in $\msun$ yr$^{-1}$, $\scl$ is the surface density 
of the clump 
in g cm$^{-2}$
in which the stars are forming, and
the parameter $j$ is determined by the density profile of the core; we
follow \citet{mckee03} in setting $j=\frac 12$. \citet{tan04} suggested that
the coefficient in the accretion rate in equation (\ref{eq:mdtc})
be increased by a factor $\sim 2.6$ in order
to allow for infall at the beginning of the inside-out collapse. We retain the value
given by \cite{mckee03}, but note that the overall normalization of the accretion
rate remains uncertain.

\item[iii)] A simplified model to represent
competitive accretion (CA, \citealt{bonnell97,bonnell01}),
\beq
\dot m =\mdone\left(\frac{m}{\mf}\right)^{2/3}\mf,
\label{eq:mdca}
\eeq
where $\mdone$ is the accretion rate for $m=\mf=1$. Note that this accretion rate
has the property that stars of all masses form in the same amount of time.

\end{itemize}

All these forms for the accretion history can be summarized in the expression
\beq
\dot m=\mdone\left(\frac{m}{\mf}\right)^j\mf^\jf, 
\label{eq:md}
\eeq
where the model exponents are given by:
\beqa
{\rm IS}: & & j=\jf=0\\
{\rm TC}: & & j=\frac{1}{2}, \jf =\frac{3}{4}\\
{\rm CA}: & & j=\frac{2}{3}, \jf=1.
\eeqa

We also considered a two-component turbulent core model (2CTC), a blend
of the IS and TC models:
\beq
\dot m=\mds\left[1+\rmd^2\left(\frac{m}{\mf}\right)^{2j}\mf^{3/2}\right]^{1/2},
\label{eq:mdtctc}
\eeq
where $\calr_{\dot m}=\dot m_{\rm TC}/ \dot m_{\rm IS}$; we adopt
$\calr_{\dot m}=3.6$ as in Paper I.
This model is similar to the TNT model of \citet{myers92}.
There is some evidence that a two-component accretion
model may be more physical in the CA case as well (i.e., a 2CCA
model). \citet{smith09} show that, at least
initially, competitively accreting protostars are surrounded by a small bound
envelope. This suggests that at early times the protostars may accrete
via a Shu-like constant accretion rate. However, the authors find
that the envelope mass is not well correlated with the
final mass of the star, which they suggest indicates that the CA phase dominates. 
For comparison, we define a 2CCA model with:
\beq
\dot m=\mds\left[1+\rmdca^2\left(\frac{m}{\mf}\right)^{4/3}\mf^2\right]^{1/2},
\label{eq:mdcaca}
\eeq
where 
\beq
\rmdca\equiv \frac{\mdca}{\mds}.
\eeq
According to \citet{smith09}, accretion of the core envelope should
contribute less than half of the final mass. Therefore, we adopt 
$\rmdca = 3.2$, which is determined by assuming that a
star of average mass accretes half its mass from its envelope.

In the fiducial cases above, the accretion rates increase
monotonically over the protostellar lifetime. 
For the single-component models,
the time to form a
star, $\tf$ is a power-law function of the final mass,
\beq
\tf=\tfo \mf^{1-\jf},
\eeq
where
\beq
\tfo = \frac{1}{(1-j)\mdone}
\eeq
is the time to form a star with $m=1$.
In reality, we
expect accretion rates to gradually taper off
(e.g., \citealp{myers98,myers10}), 
at least for the IS and TC models.
Tapered accretion is supported observationally since Class I sources are not significantly brighter than Class 0 sources, and in some cases, the most luminous young source is actually Class 0 (e.g., \citealt{evans09}).
 Class I sources are also twice as numerous, suggesting that high accretion rates, which depend upon the presence of dense infalling gas, cannot be sustained throughout the majority of the formation time \citep{enoch09}.
In competitive accretion models, the decline in accretion rates is
assumed to be abrupt, so untapered models are likely to be the
best approximation for the CA case.
If we assume that the decline is linear in time, the tapered accretion
case can be written:
\beq
\dot m = \mdo(m, \mf) \left[1-\left({{t}\over{\tf}}\right) \right],
\eeq
where $\mdo (m,\mf)$ is the untapered accretion rate for $m=\mf = 1~\msun$.
We can then express the accretion rate for both the tapered and untapered cases as
\beq
\dot m=\mdoo\left(\frac{m}{\mf}\right)^j\mf^\jf\left[1-\delta_{n1}\left(\frac{m}{\mf}\right)^{1-j}
\right]^{1/2}
\label{eq:mdt}
\eeq
where $\mdoo$ is the untapered accretion rate for $m=\mf=1$ and
\beq
\delta_{n1}=\left\{ \begin{array} {r@{\quad\quad} l}0 & {\rm untapered},\, n=0,\\
	1 & \mbox{tapered},\, n=1, \end{array}  \right.
\eeq
The star formation time for 
single-component models can then written as:
\beq
\tf=\tfo \mf^{1-\jf} (1+\delta_{n1}).
\eeq

In the case of accelerating star formation we assume a birthrate that
increases exponentially with time, i.e.,
\beq
\dot {\cal N}_* \propto e^{(t-t_m)/\tau}, 
\eeq
where $t_m$ is the age of a star with mass $m$ and final mass $\mf$ and
$\tau$ is the characteristic acceleration time. For an accelerating star
formation rate, the mean formation time is no longer equal to the
observed star formation time. Instead,
\beqa
\avg{\tf}_{\rm obs}& =&
\frac{\avg{t_{\rm II}}\avg{1-e^{-\tf/\tau}}}{\avg{e^{\tf/\tau}}(1-e^{-t_{\rm
	II}/\tau})} \\
& \simeq& \left[\frac{\avg{t_{\rm II}}}{\tau(1-e^{-\avg{t_{\rm
	  II}}/\tau})} \right] \avg{\tf} \label{tfacc},
\eeqa
where $t_{\rm II}$ is the lifetime of Class II sources. For $t_{\rm
II} = 2$ Myr and $\tau = 1$ Myr, the 
actual mean star formation time of the
accelerating cases, $\avg{\tf}$, is a factor $\simeq 2.3$ smaller than the value
inferred from observations of the numbers of protostars and young stellar objects,
$\avg{\tf}_{\rm obs}$.

\section{The Protostellar Luminosity Function (PLF)} \label{s.plf}

\subsection{Definition}

Our objective is to determine the protostellar luminosity function, $\Ppl$,
where $\Ppl d\ln L$ is the fraction of protostars that have luminosities in the 
range $dL$. The protostellar mass 
and accretion rate are related to the accretion luminosity through equation (\ref{eq:lacc}),
which gives $L\propto m\dot m/r$. Unfortunately this relation is complicated
by the fact that
the protostellar radius is a function of both the mass and the accretion rate,
$r=\rmmd$, which does not have a simple analytic representation \citep{stahler88,hartmann97}.
We use the routine described by \citet{offner09}, which agrees with
the results of \citet{stahler88}, to determine the protostellar radius
(see \S \ref{1dmodel}).
This model exhibits evolution similar to the model with shock boundary
conditions in \citet{hosokawa11}. Consequently, the subsequent
evolution of the stars will have ages that are consistent with
previously calculated non-accreting isochrones.

We proceed by using equation (\ref{eq:md}) to determine $r(m,\mf)$ from
$r(m,\dot m)$,
and then solving the accretion luminosity equation (\ref{eq:lacc}) for
$\mlmf$.
Since there are two independent variables,
we use bi-variate distribution functions: The fraction of protostars in the
luminosity range $dL$ and final mass range $d\mf$ is the same as the fraction
protostars with masses in the range $dm$ and final mass range $d\mf$:
\beq
\Ppt(L,\mf) d\ln L\, d\ln \mf=\ppt(m,\mf) d\ln m\, d\ln\mf, \nonumber
\label{eq:Ppt}
\eeq
where $\ppt$ was determined in Paper I (see eq. \ref{eq:ppt}).
The PLF is then obtained by integrating $\Ppt(L,\mf)$ 
over $\mf$,
\beqa
\Ppl&=&\int d\ln \mf \Ppt(L,\mf),
\label{eq:pplis1}\\
&=&\int_{\mfl(L)}^\mup d\ln \mf\,\frac{\ppt[m(L),\mf]}{\displaystyle \left|\ppbyp{\ln L}{\ln m}\right|}.
\eeqa
In the Isothermal Sphere case, the accretion rates are independent
of the final mass. Consequently, the PLF reduces to:
\beq
\label{eq:pplis2}
\Ppl  \rightarrow \frac{m(L)}{\avg{\mf}}\int_\mfl^\mup d\ln\mf\;\frac{\psi(\mf)}{\left|1-\ppbyp{\ln r}{\ln m}\right|}.
\label{eq:pplis3}
\eeq

\subsection{The Mean and Standard Deviation of the Luminosity}\label{s.meanl}

In general, the mean luminosity can be evaluated as $\avg{L}=\int L\Ppl d\ln L$. However,
it is more instructive to return to the bi-variate luminosity function
and define the mean accretion luminosity:
(\ref{eq:Ppt}):
\beqa
\avg{L_{\rm acc}}&=&\int_\ml^\mup d\ln\mf\int_0^\mf d\ln m\, L_{\rm acc}\ppt,\\ 
&=& \frac{1}{\avg{t_f}}\int_\ml^\mup d\ln\mf
\psi(\mf)\left[\frac{\facc G\mf^2}{2\bar r(\mf)}\right] \label{eq:binding},
\eeqa
where equation (\ref{eq:Ppt}) enabled us to eliminate the dependence
on $\dot m$ for steady star formation and where
\beq
\frac{1}{\bar r(\mf)}\equiv \frac{2}{\mf^2}\int_0^\mf \frac{mdm}{r(m,\mf)}
\eeq
is the mass-averaged harmonic mean radius.
Note that equation (\ref{eq:binding}) is only valid in the case of steady
star formation, where the accretion luminosity is proportional
to the binding energy of the stars formed.
In order to evaluate the mean luminosity, we take advantage of
the fact that the stellar radius depends primarily on the instantaneous
mass $m$ and only weakly on the final mass $\mf$ in order to 
define a mean radius in terms of the current mass $m$,
\begin{equation}\label{eq:barr2}
\frac{1}{\bar r(m)} = \frac{2}{\mup^2-m^2}\int_{m_{f,l}}^{\mup}\frac{\mf d\mf}{r(m, \mf)}.
\end{equation}
This approximation is reasonably accurate for the value of
the upper mass limit
we consider here ($\mup=3$); for larger values of $\mup$, it would be
preferable to include a weighting with the IMF.

For protostars with $\mf \la 1~\msun$, the stellar luminosity, i.e., the
luminosity due to nuclear burning
and Kelvin-Helmholtz contraction, makes only a small
contribution to the total luminosity, so that $\avg{L} \simeq
\avg{L_{\rm acc}}$.
However, for more evolved protostars beginning late in the Class I stage,
accretion diminishes and no longer dominates the
total luminosity \citep{white04}. For the young, low-mass, embedded sources
we consider here, 
the accretion luminosity is much greater than the nuclear luminosity
provided $m \la 1.5 \msun$. For such stars with tapered
accretion, the accretion
luminosity dominates during the majority of the evolution, i.e., while
$m \le 0.95~\mf$. 
For stars with $m > 1.5 \msun$, the nuclear
luminosity becomes important and the accretion luminosity is not a
good approximation of the total. However, these stars comprise only a
small part of the total sample. For example, the maximum expected
protostellar mass, $\mmax = 2.3 \msun$, for a cluster with $\mup
= 3 \msun$ and $\caln_p=112$
(see Figure 7 in Paper I) in the CA case corresponds to a
maximum ratio of
$L_{\rm acc}/ L_{\rm tot} = 0.65$. In the fiducial IS case, $m_{\rm
max}= 2.9 \msun$ yields $L_{\rm acc}/ L_{\rm tot} = 0.16$,
which reflects both the lower accretion rate of the IS model 
and the higher maximum mass for a cluster with a fixed number of protostars in this model.

The standard deviation is a useful metric for characterizing the
breadth of the protostellar
luminosity distribution. Since the luminosity spread may encompass several
orders of magnitude, we adopt the standard deviation of the logarithm
of the protostellar luminosity: 
\beqa
\sigma({\rm log} L)&=& \left[ \int ({\rm log}\,L)^2 \Ppl d\ln L \right.
\nonumber\\ 
& & - \left. \left(\int{\rm log}\,L\; \Ppl d\ln L \right)^2 \right]^{1/2}.
\eeqa
This metric has the advantage that it is
dimensionless and, like the
mean, it can be computed
solely in terms of the protostellar mass function if $L(m,\mf)$ is
provided. The ratio of the median to the mean, also a useful
dimensionless number,  measures the skewness of
the distribution.

\subsection{Stellar Evolution Model} \label{1dmodel}

\begin{figure}
\epsscale{1.2}
\plotone{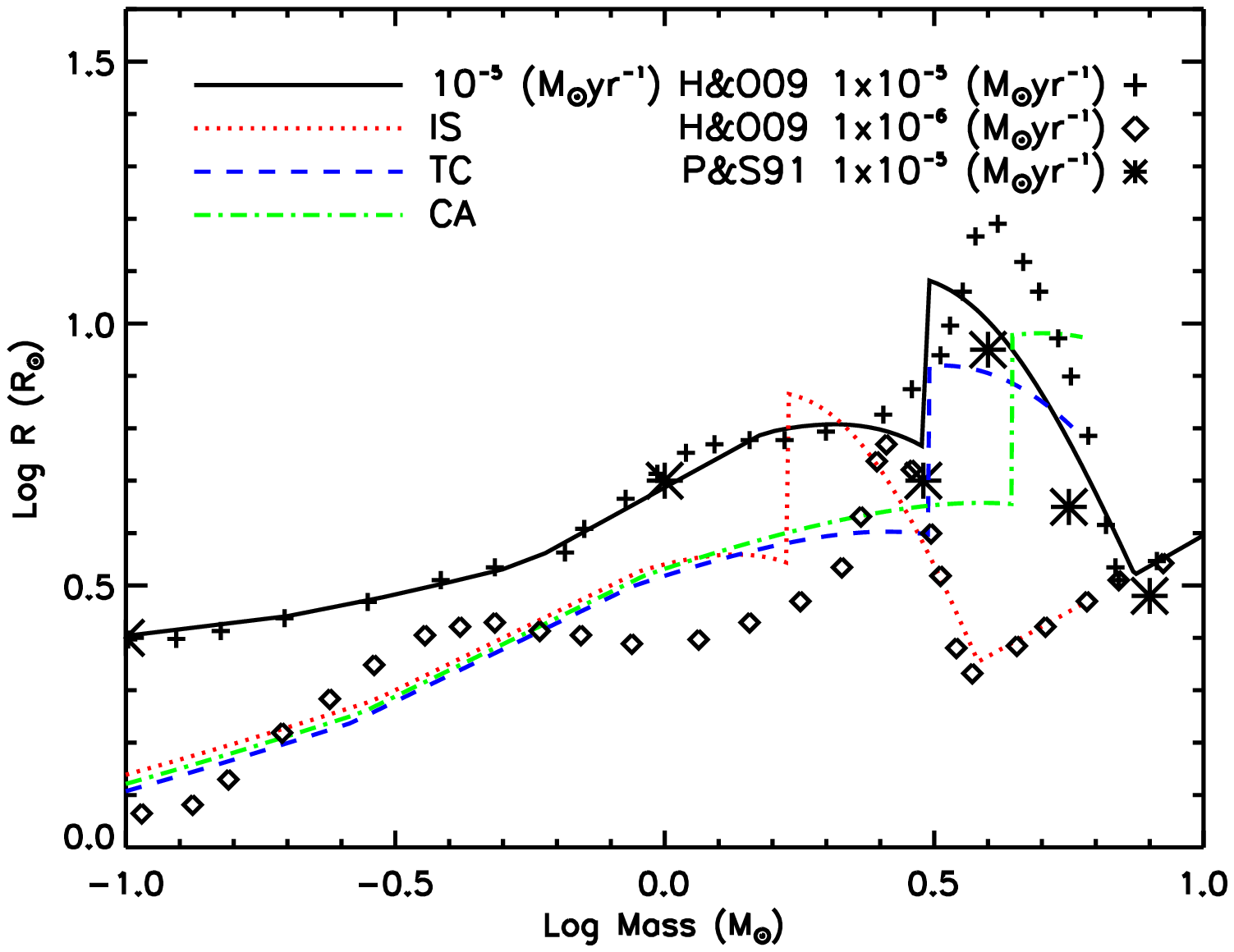}
\caption{ \label{rvsm} 
Stellar radius versus mass for a constant accretion rate of
$10^{-5}\,\msun\, \rm{yr}^{-1}$ (solid), the IS model ($\mdot=9\times 10^{-7}\,
\msun\, \rm{yr}^{-1}$, dotted), the TC
model (dashed), and the CA model (dot-dashed) where $\avg{\tf} = 0.44$ Myr
and $\mf=10 \,\msun$. 
For comparison,
models calculated by \citet{hosokawa09} for
constant accretion rates of $10^{-5}\, \msun \,\rm{yr}^{-1}$ and $10^{-6}\,
\msun \,\rm{yr}^{-1}$  are shown by the 
plus signs
and diamonds,
respectively. Stars indicate the \citet{palla91} model for a constant
accretion rate of $10^{-5} \,\msun\,
\rm{yr}^{-1}$. }
\end{figure}

In order to determine the PLF predicted by a given theory
of star formation,
we must first construct a comprehensive luminosity model
that takes into account both stellar evolution and the accretion
history. 
We adopt the one-zone protostellar evolution model developed by
\citet{tan04}, which is described in detail in
\citet{offner09}. 
This model has been calibrated to replicate the more 
detailed calculations of \citet{palla91, palla92} and 
updated to reflect the recent recommendations of
\citet{hosokawa09}. It corresponds to the ``hot'' stellar evolution
model described in \citet{hosokawa11}, where an accretion flow
directly hits the stellar surface and forms an accretion
shock front. Although the gas may be channeled through a disk,
the model assumes that it is sufficiently thick such that the accretion column covers most of the stellar surface. 
In practice, the initial physical conditions and accretion flow
properties are not well known, introducing uncertainties in the
stellar radius of a factor of $\sim 3$ for $m \le 0.2 \msun$.
(e.g., \citealt{hosokawa11}).
Figure \ref{rvsm} shows the stellar radius predicted by this model as
a function of stellar mass for different accretion histories. 
For $\dot
m = 10^{-5}\,\msun\, \rm{yr}^{-1}$, the
model has a maximum disagreement with \citet{hosokawa09} of $\sim20$\% when $m 
\leq 3\msun$ and generally agrees
within 5\%. The model has a maximum
disagreement of a factor of $\sim$2
 at $m=1.6\msun$ for $\dot m =
10^{-6}\,\msun\, \rm{yr}^{-1}$, but it generally is within a factor of 1.2
\citet{hosokawa09}. 
Although our sub-grid model has been thoroughly described in \citet{offner09}, 
we give a brief summary here. 

The model treats the protostar as a polytrope and follows the
stellar contraction 
by enforcing energy conservation. The model is characterized by six 
stages, culminating in the arrival of the protostar on the zero age 
main sequence. In the``pre-collapse'' stage, the collapsing gas
densities 
are sub-stellar.  In the second ``no burning stage'', the densities
are sufficient for the dissociation of H$_2$, but are too low 
for deuterium burning. In the following ``core deuterium stage at fixed 
$T_c$,'' deuterium burning ignites in the core.  Once the supply of
deuterium is 
depleted, the core temperature rises and the protostar enters the 
``core deuterium burning at variable $T_c$.''  
Eventually high temperatures in the core terminate convection and halt
the 
flow of deuterium into the core, whereafter the protostar enters
the 
``shell deuterium burning'' stage. Finally, the central temperature 
reaches $10^7$ K, and the star moves onto the main sequence. 
In calculating the total luminous output, the zero-age main sequence
luminosity serves as an approximation of the interior luminosity,
which results from the combination of deuterium burning and 
Kelvin-Helmholz contraction. 
We evolve 
this model in combination with the accretion rates specified by the
star 
formation models and generate luminosity tables as a function 
of the instantaneous and final stellar masses for each case with both untapered and tapered 
accretion. 
For an accelerating star formation rate, we assume that
the 
physical parameters are set by the initial conditions of the collapse 
and hence do not themselves accelerate. Thus, for the untapered 
accelerating and tapered accelerating star formation cases, 
we use the untapered and tapered luminosity tables, respectively.

We use these tabulated values of $L(m,\mf)$ to obtain the mean and
standard deviation of the luminosity directly. However, in the model, $L$ and $r$
undergo discontinuous jumps as the deuterium state changes. This is
problematic in calculating the PLF, which requires the derivative of $L$. To circumvent
this issue, we use a polynomial fit to equation (\ref{eq:barr2}). 
We re-normalize the result using the directly derived value of $\avg{L}$.
This strategy simplifies the integral while preserving the trends in
the sub-grid model and eliminating the weak dependence of
$L$ on $\mf$.

\subsection{Episodic Accretion}
\label{sec:epi}

Observations of high-luminosity variable sources, such as the
prototype FU Ori, suggest that protostars undergo periods of high
accretion (e.g., \citealt{hartmann96}).
In the extreme case,
protostars may spend most of their life in a low-luminosity,
low-accretion phase and accrete nearly all their mass during short, intense
accretion bursts. However, only 
a total of 18 
bursting sources have been identified within 1 kpc of the Sun
\citep{greene08}. These sources have luminosities in the range
20-550 $\lsun$, corresponding to accretion rates of $10^{-5}-10^{-4}
\msun$ yr$^{-1}$. 
The total mass that can be accreted in such
bursts is limited by the known star formation rate. Following
\citet{mckee10b}, let $f_{\rm epi}$
be the fraction of mass accreted during episodic accretion periods:
\begin{equation}
f_{\rm epi} = {{\dot m_{\rm epi} \Delta t_{\rm epi}}\over{\avg{m_f}}},
\end{equation}
where $\Delta t_{\rm epi}$ is the total time spent in the bursting
state and $\avg{m_f} =0.5$ is the mean stellar mass for a typical
IMF. The time spent in the high accretion state can be expressed as: 
\begin{equation} 
\Delta t_{\rm epi} = {{\caln_{\rm p, epi}}\over{\nds}},
\end{equation}
where $\caln_{\rm p, epi} \simeq 20$ is the number of protostars
accreting in this state within 1 kpc of the Sun \citep{greene08} and the star formation
rate, $\nds$, is 0.016 stars/yr within 1 kpc \citep{fuchs09}.
This gives $\Delta t_{\rm epi} \simeq 1200$ yr and $f_{\rm epi}
\simeq$ 0.25 for an accretion rate
of $10^{-4}~\msun $ yr$^{-1}$. If the protostellar lifetime is 0.5
Myr, then the duty cycle must be $\sim$ 0.2\%.

This estimate suggests that our
sample of 112 protostars is unlikely 
to contain any bursting sources. For our
observational comparison, we adopt an episodic accretion factor, $f_{\rm epi} =0.25$,
to reflect the amount of mass accreted during outbursts. 
If no protostars are currently undergoing bursts, then the
observed sample will have a mean luminosity that is 75\% of the true
time-averaged mean. To account for this, we reduce our model
mean luminosity by a factor of $1-f_{\rm epi} = 0.75$. 
In other words, we adopt an effective value for $\facc$ in equation (\ref{eq:lacc}) of
\beq
f_{\rm acc,\,eff}=\facc(1-f_{\rm epi})=0.56.
\eeq
Note that
the \citet{dunham10} episodic accretion model corresponds to $f_{\rm
epi}\simeq 0.7$ for a star with $\mf=0.35$.
For $\facc=0.75$, this corresponds to $f_{\rm acc,\,eff}=0.23.$
An alternative way of correcting for episodic accretion assumes that
the protostars accrete at their normal rate for a time 
$t_{f,\,\rm non-episodic}$
and then accrete 
via bursts for a brief additional time.
In this case, the luminosity of the protostars would be unchanged and
instead the formation time would be correspondingly shorter. We adopt
the former definition of $f_{\rm epi}$ since it seems more plausible
and is consistent with simulations exhibiting episodic accretion
(e.g., \citealt{vorobyov05}).

\section{COMPARISON WITH OBSERVATIONS} \label{s.comp}
\subsection{Overview of the Observational Data}

Throughout this section, we compare to the Class 0 and Class I source 
luminosities of Serpens, Ophiuchus, Perseus, Lupus and Chameleon from  
\citet{enoch09} and \citet{evans09}. This is the same sample used
in \citet{dunham10}. It has been corrected for localized extinction and
carefully analyzed to remove non-protostars and non-embedded sources (Mike Dunham, private communication).
The data are comprised of 39  Class 0 and 73  
Class I objects for a total sample size of 112. This includes only two sources from
Chameleon II and one source from Lupus, such that the
sample essentially reflects the properties of Serpens, Ophiuchus, and Perseus. The Class 0 and Class
I sources have a 0.50 likelihood of being drawn from the same
population, which suggests that the two groups are quite similar. It is
unclear if this is due to similar accretion rates or source mixing resulting
from inclination effects (e.g., \citealt{robit06}).

As discussed 
by \citet{enoch09} and \citet{dunham08}, the bolometric luminosities have an uncertainty 
of $\sim$ 50\% due to saturation of the 160 $\mu$m band.  
\citet{enoch09} estimate that $\sim$ 10\%  error arises from the SED 
fitting process and 15-25\% due to finite sampling errors. 
Although the data we use have improved 350 $\mu$m
observations and more careful source selection, uncertainty arises from a number of
factors and is not well constrained.
Following \citet{enoch09}, we adopt a 50\% upper uncertainty in the bolometric
luminosities, and we use the non-extinction corrected bolometric luminosities,
which are naturally underestimates, to set a lower error limit
(Melissa Enoch
and Mike Dunham, private communication). The extinction--corrected mean and median bolometric
luminosities are then: 
\beqa
\avg{L_{\rm obs}}& =& 5.3^{+2.6}_{-1.9}~\lsun, \\
L_{\rm obs, med}& =& 1.5^{+0.7}_{-0.4}~\lsun.
\eeqa 
Note that the mean is very sensitive to the
most luminous sources. 
The extinction corrected sample of
\citet{enoch09}, which is derived from a nearly identical source list, has a mean
of $3.5~\lsun$,which falls within the error of our adopted mean. 
The disagreement arises from a combination of improved 350 $\mu$m data
used in the \citet{evans09} analysis and different treatments of the IRAS
fluxes. 
In either case, the luminosity of a few of the brightest sources may have
been overestimated by as much as a factor of two if the
24$\mu$m flux was omitted due to saturation 
(Mike Dunham, private communication).
Consequently, the upper uncertainty should be
considered an upper limit on the luminosity rather than a one-sigma
error estimate. Since the
standard deviation is also sensitive to the distribution outliers, we
use the log of the standard deviation for comparison: 
\begin{equation}
\sigma({\rm log} L_{\rm obs}) = 0.7^{+0.2}_{-0.1}~\rm{dex}. 
\end{equation}
A second
useful non-dimensional quantity is the ratio of the median luminosity
to the mean: 
\begin{equation}
L_{\rm med}/\avg{L} =0.3^{+0.2}_{-0.1}. 
\end{equation}
However, the associated uncertainties
of this ratio 
are somewhat uncertain since the median and mean errors are correlated.

\begin{figure}
\epsscale{1.2}
\plotone{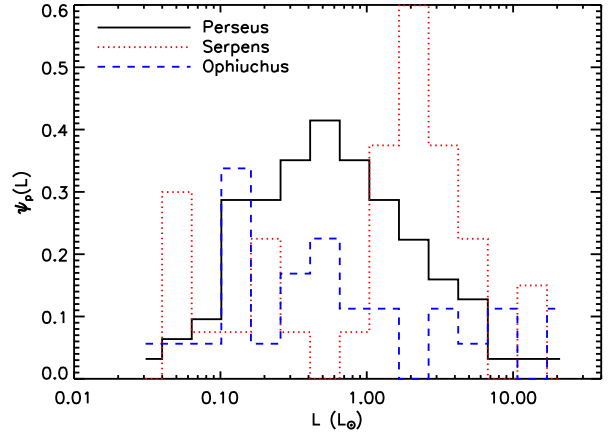}
\caption{ \label{plfobs} 
The PLF, $\Ppl$, of the observed protostellar luminosities for
Perseus, Serpens, and Ophiuchus \citep{evans09}.}
\end{figure}

In this sample, the mean luminosities of the individual observed star forming 
regions and their shapes are similar, even though the number of
sources in each is different. The PLFs of Perseus, Serpens, and
Ophiuchus are shown in Figure \ref{plfobs}, where the regions have mean
luminosities 5.2 $\lsun$, 5.0 $\lsun$, and 6.2 $\lsun$, respectively.
A Kolmogorov-Smirnov (K-S) test indicates that Perseus and Serpens have
a 0.39 likelihood of of being drawn from the same parent distribution, 
while Perseus and Ophiuchus have a 0.45 likelihood and Ophiuchus and 
Serpens have a 0.86 likelihood.
This high level of agreement, despite the apparent dissimilarity of the
distributions shown by Figure \ref{plfobs}, is partially due to the small
number of protostars ($\sim$ 20) in Serpens and Ophiuchus. 
While comparing our models with the individual regions
would be ideal, small statistics requires that we use the total
protostellar distribution over all the clouds.

To derive the model distribution, we require the mass of the most massive star
forming in the cluster, $m_u$. The brightest Class II source in
\citet{evans09} suggests an upper mass of $\sim 3~\msun$. Since the
number of Class
II sources is several times larger than the sample of Class I and
Class 0 sources, this should statistically be a good upper limit. However,
statistical variations or changes in the star
formation rate may impact the upper mass of the currently forming stars.
Inspection of the most luminous Class I source also suggests an upper
mass of $3 \msun$, assuming
that its luminosity is dominated by stellar radiation rather than
accretion. This latter
estimate is also very uncertain since we can't discount that the
source is undergoing an outburst and its luminosity is dominated by
accretion. However, we can be fairly confident that $\mup=3~\msun$ is
a good upper limit.

The accretion timescale of the models is constrained by the observed
protostellar lifetime, i.e.,  how long protostars spend in the main
accretion phase. Both Stage 0 ($M_{\rm env} > m$) and Stage I
($m > M_{\rm env}> 0.1 \msun$) protostars experience significant accretion, since a
significant fraction of the total gas is contained in the envelope \citep{crapsi08}. 
These stages roughly correspond to Class 0 and Class I provided that the Class I
envelope mass exceeds 0.1 $\msun$. \citet{enoch09} estimated 0.1 $\pm0.02$ Myr
for the Class 0 lifetime.
This is longer than the Class 0 lifetime of $1-3\times 10^4$ yr
reported by \citet{andre00}, because 
these authors
base their lifetime solely on data from Ophiuchus, which is now
recognized to have a deficit of Class 0 objects compared to other star
forming regions (\citet{enoch09}, who also include one more Class 0 object than \citet{andre00} in their Ophiuchus sample.)
For local star forming regions, \citet{evans09} report an average
combined Class 0 and I lifetime of 0.44 Myr.
The uncertainty in this estimate
arises in part from statistics and Class I/II source confusion, but it is
dominated by the uncertainty in the disk lifetime ($2 \pm 1$
Myr), which is necessary to calibrate the ages.
Altogether, the mean protostellar lifetime is
$\avg{ \tf } = 0.44 \pm 0.22$ Myr in the \citet{evans09}
sample.  Their estimates of the mean protostellar lifetimes 
in individual clouds, 
Ophiuchus (0.30 Myr), Serpens (0.46 Myr), and Perseus (0.62 Myr), are
within the errors of the overall mean.
This lifetime is significantly longer than
previous estimates, which have adopted either the Class 0 lifetime
(e.g., \citealt{kenyon90}) or the core free-fall time ($\sim$0.1
Myr for a 0.5$\msun$ star; e.g., \citealt{myers98, young05}). Shorter
lifetimes exacerbate the luminosity problem. As discussed by
\citet{mckee10b}, one possible solution is ``slow accretion,'' in which
accretion rates are reduced by protostellar outflows or
lengthened disk lifetimes. Estimated protostellar lifetimes on the order of $0.5$ Myr
lend credence to a slow accretion picture.

Some studies of Class I protostars have found that stellar luminosity
dominates the total luminosity, suggesting that accretion has already
diminished \citet{muzerolle98, white04, connelley10}. 
However, in-depth study often reveals that sources classifed as Class
I are actually edge-on or heavily obscured Class II sources
\citep{white04,kempen09, heiderman10}.
The observational sample here uses a minimum gas mass criteria of $0.1\msun$ for
inclusion
in the sample, which is an indicator that accretion is still
underway. However, 
the authors 
do not probe for tracers of dense, warm
 gas, such as emission from higher transtions of HCO$^+$ and
 C$^{18}$O, which
more conclusively separates embedded protostars from obscured Class IIs \citet{kempen09, heiderman10}.

\begin{figure}
\epsscale{0.95}
\plotone{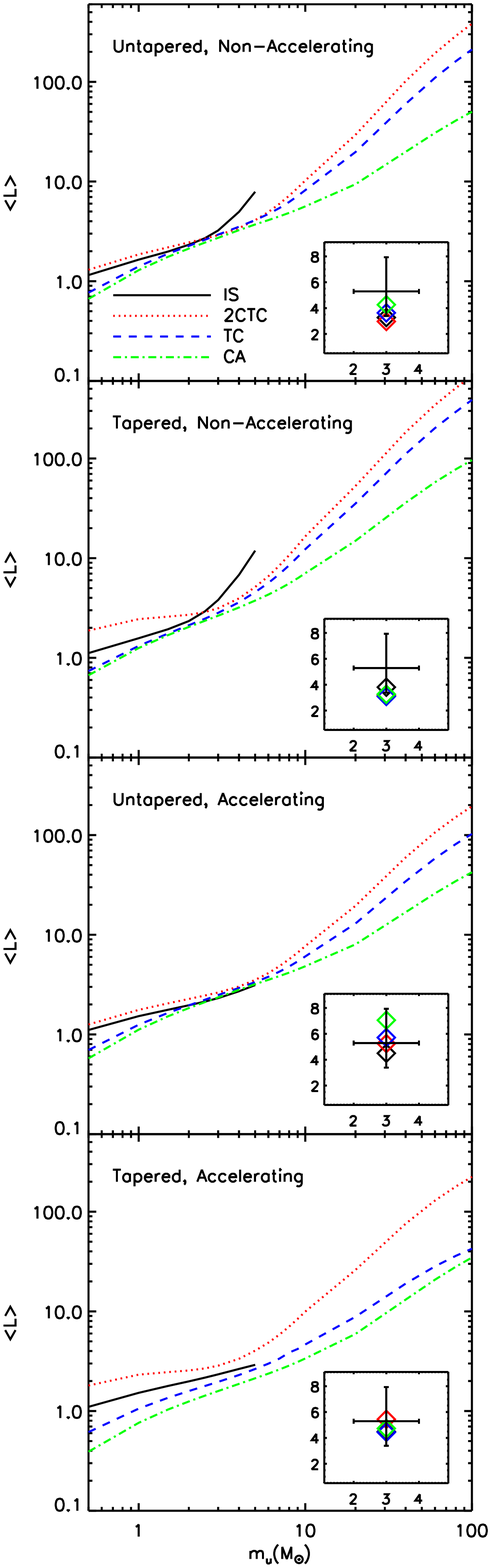}
\caption{ \label{meanlvsmu} 
The mean total protostellar luminosity versus
the upper protostellar mass, $\mup$, for each of the models with
$\avg{\tf} = 0.44$ Myr. For the tapered cases, $n=1$, and
for the accelerating cases,
$\tau=1$ Myr. The inset shows the mean observed luminosity from \citet{evans09} with error
bars representing the uncertainty in the measurement and $\mup$. The diamonds display the
values in 
Table 1.
These values assume $L_{\rm min} = 0.05~\lsun$ and
$\avg{\tf}_{\rm obs}=0.44$ Myr, which corresponds to $\avg{\tf}=0.24$ Myr for the
accelerating cases.}
\end{figure}

\begin{figure}
\epsscale{1.2}
\plotone{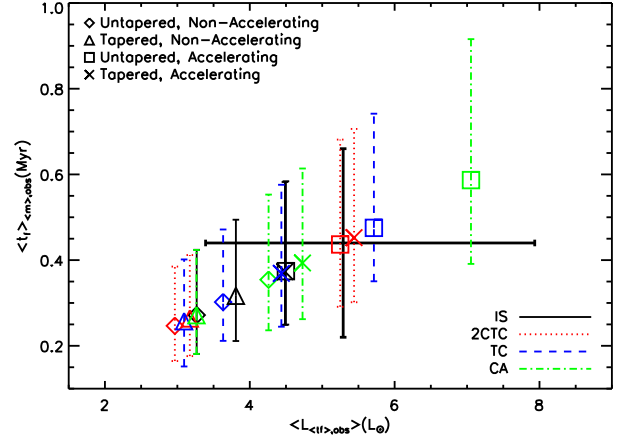}
\caption{ \label{tffig} 
Protostellar lifetime estimated using the observed mean
luminosity from the \citet{evans09} data as a function of the
mean luminosity from the models. The error bars on the model lifetimes
derive from the observal luminosity uncertainty.
The two observational results with uncertainty are shown by the thick set of solid error bars.}
\end{figure}

\begin{figure}
\epsscale{2}
\plottwo{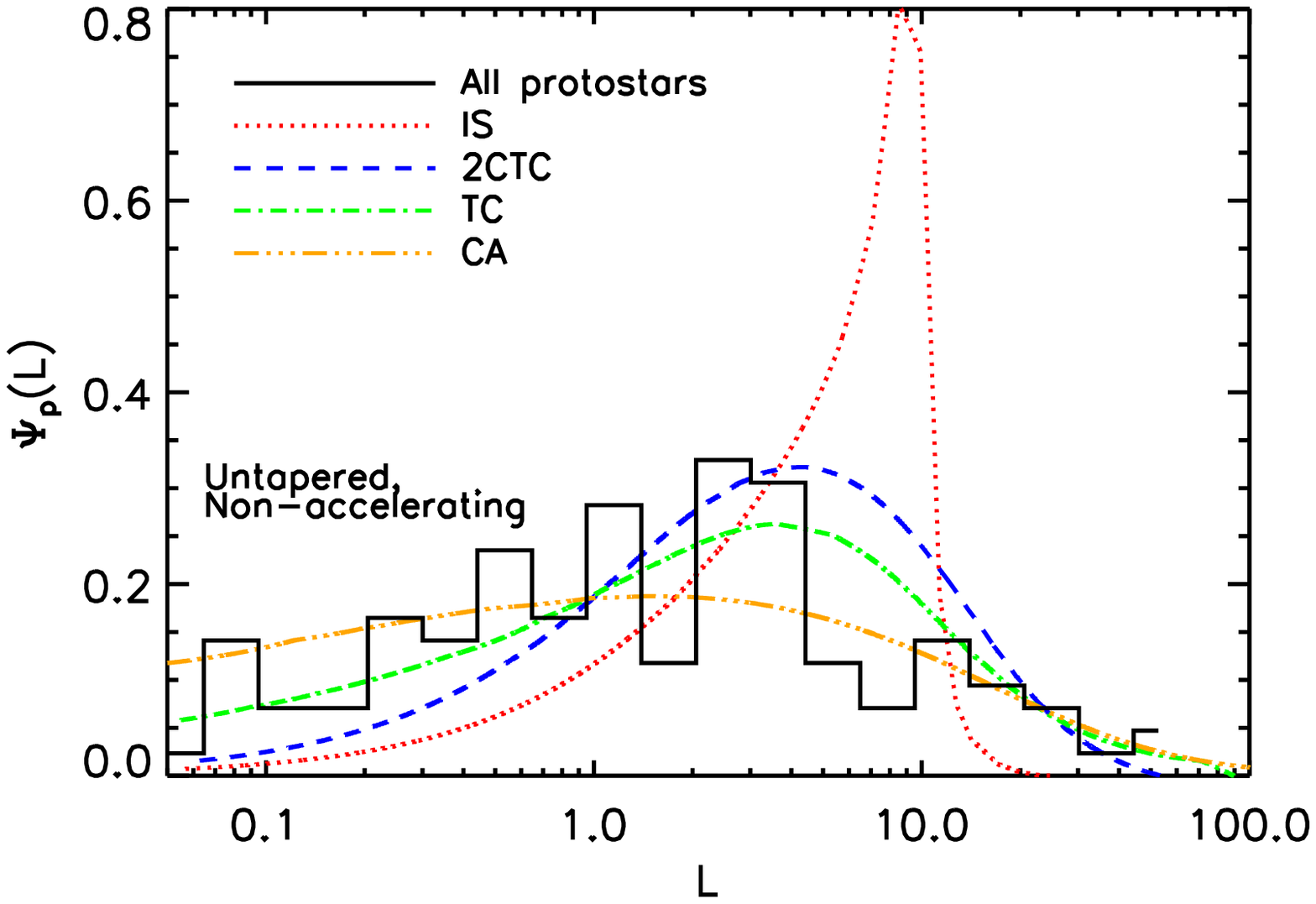}{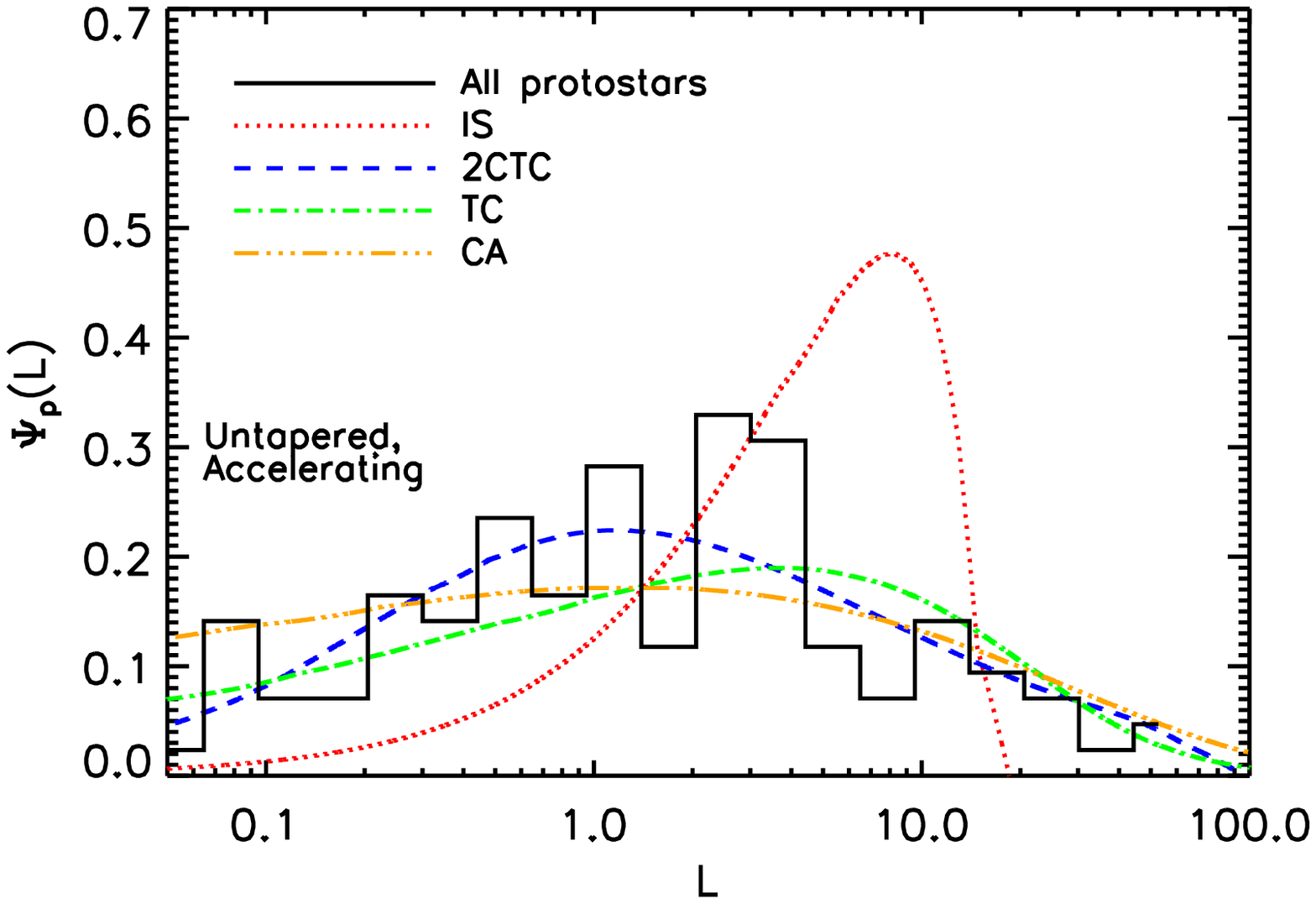}
\caption{ \label{plfshapes} 
The PLF for $m_u = 3\msun$ for untapered, non-accelerating star
formation (top) and untapered, accelerating star formation
with $\tau = 1$ Myr (bottom). The observed PLF \citep{evans09} is
plotted for comparison. 
Note that the PLF shape is derived assuming that the accretion
luminosity dominates the total.}
\end{figure}

\begin{figure}
\epsscale{2.0}
\plottwo{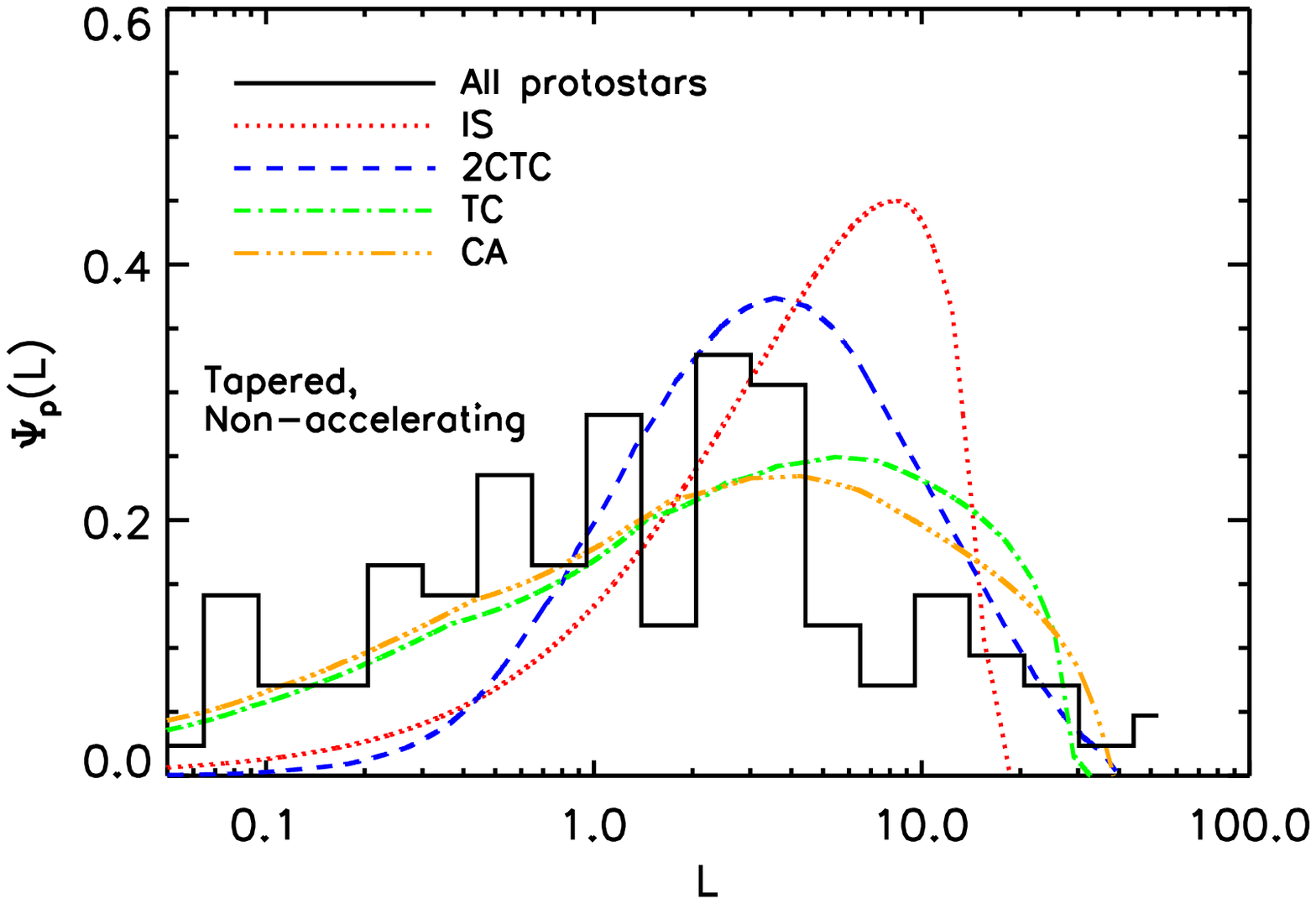}{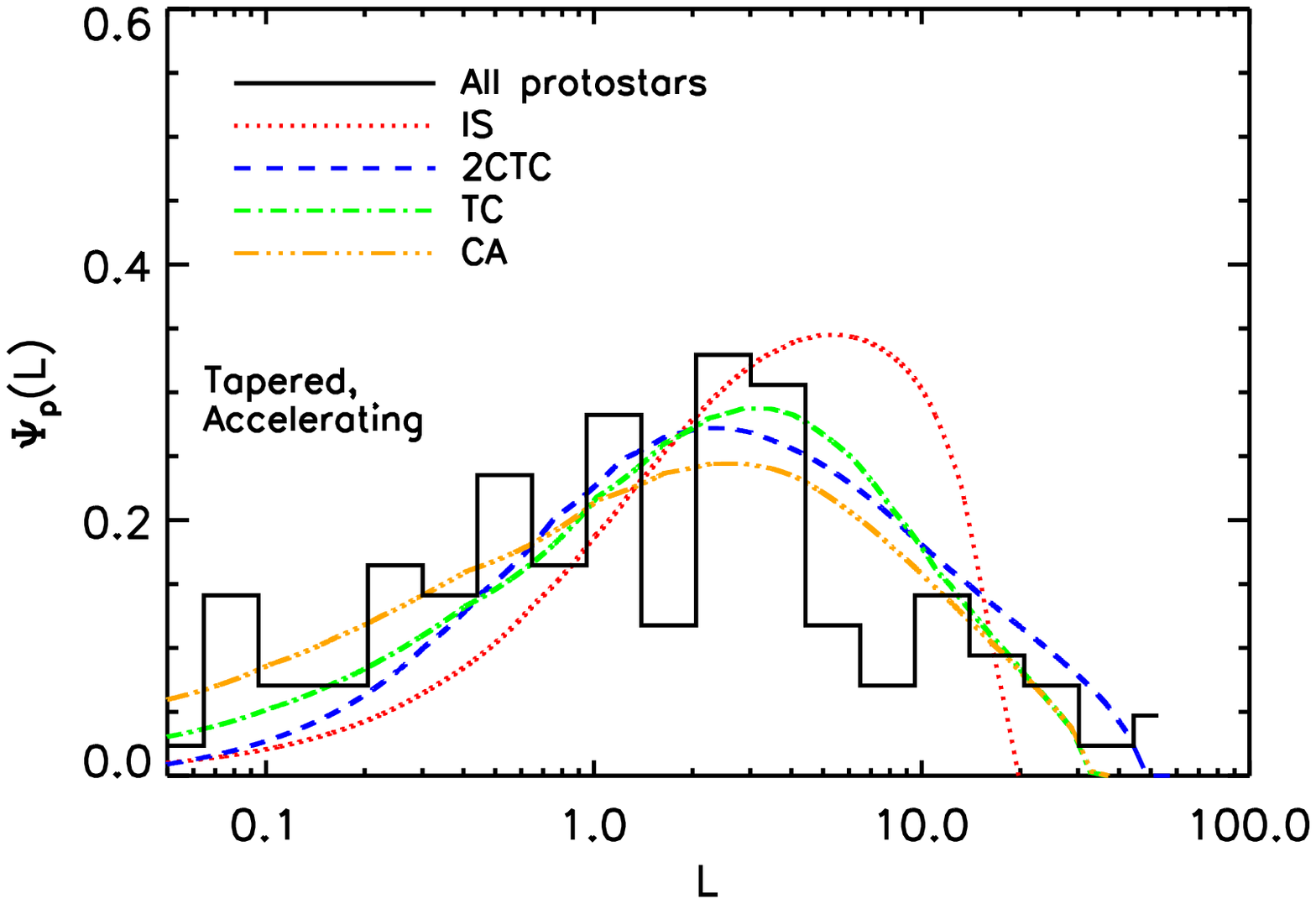}
\caption{ \label{plfshapes2} 
The PLF for $m_u = 3\msun$ with tapered accretion rates. The observed
PLF \citep{evans09} is plotted for comparison.
Note that the PLF shape is derived assuming that the accretion
luminosity dominates the total.
}
\end{figure}

\subsection{The Mean Luminosity, $\avg{L}$}\label{meanl}

The mean of the luminosity
distribution serves as a
simple one-dimensional statistic with which to compare the 
models and observations directly. 
In Figure \ref{meanlvsmu}, we plot the mean 
luminosity as a function of the most massive star forming in the
cluster for each of the accretion models. We truncate the IS model 
at $5\msun$ since it is unrealistic
for describing high-mass star formation.
In Table \ref{tablemean}, we give the model
values calculated assuming completeness 
of the extinction-corrected data
down to $0.05 L_\odot$ for $\mup=3 \msun$.
We plot these tabulated values and the 
observed mean ($\avg{L_{\rm obs}} = 5.3^{+2.6}_{-1.9}~L_\odot$) 
from \citet{evans09}  in an inset plot in Figure
\ref{meanlvsmu}. 
As with the mean protostellar mass derived in Paper I, the mean
luminosity increases strongly as a function of the upper stellar
mass. Unfortunately, for $\mup=3\msun$, the model spread is small,
making it difficult to discriminate between models. In contrast, the
model means diverge significantly for clusters with large $\mup$ in
all cases,
suggesting that
more complete observations of high-mass star-forming regions would be useful to
distinguish models based solely upon the mean luminosity. Note that
the model means depend upon the  
uncertain
star-formation timescale, so
that it is not possible to make an independent comparison of the models
and observations (see discussion in \S\ref{s.dis}). However,
better constraints on the star formation time in the
future should increase the discriminating value of the mean luminosity
in comparing models.

As shown in the plot insets, the mean observed luminosity falls above
the models in all the non-accelerating cases
cases. The untapered TC and CA means, IS tapered mean, and
all accelerating means are consistent with the observational error.\footnote{The CA luminosity distribution
extends to 
both the highest and
the lowest luminosities, so that it yields
the largest mean when a cutoff of 0.05$\lsun$ is applied and the curve
is re-normalized thus weighting the highest luminosities more heavily;
the IS case is least affected by this luminosity truncation since it
is strongly peaked around a luminosity much higher than the cutoff.}
This clearly indicates that there is no luminosity
problem in the traditional sense. The resolution is a result of the longer protostellar
lifetime adopted from \citet{evans09} and
an effective accretion efficiency, $f_{\rm acc,\,eff}=0.56$ due to
a radiative efficiency of 75\%
and allowance for episodic accretion at the level of 25\%. 
Altogether this reduces the
predicted luminosities 
for the non-accelerating cases
by a factor of $\sim 3$.
				   
\begin{deluxetable*}{lllll}
\tablewidth{0pt}
\tablecolumns{5}
\tablecaption{Mean luminosity $\avg{L}$(L$_\odot$) for $\avg{\tf}_{\rm obs}= 0.44$ Myr;
$\avg{L_{\rm obs}}= 5.3^{+2.6}_{-1.9}~L_\odot$.
\label{tablemean}}
\tablehead{ 
\multicolumn{1}{c}{} & \multicolumn{2}{c}{Non-Accelerating} & \multicolumn{2}{c}{Accelerating \tablenotemark{a}}\\ 
\colhead{Model} &\colhead{Untapered} & \colhead{Tapered\tablenotemark{b}}&\colhead{Untapered }& \colhead{Tapered\tablenotemark{b}}}
\startdata
Isothermal Sphere (IS)                    &  3.27   & 3.81  &4.50   & 4.49  \\ 
2C Turbulent Core (2CTC)\tablenotemark{c} & 2.97   & 3.17  & 5.25  & 5.44  \\ 
Turbulent Core (TC)                       &  3.63  & 3.09  & 5.70 &4.44  \\
2C Competitive Accretion (2CCA)\tablenotemark{d} &   2.91  & 2.68  &
4.97 & 4.14 \\
Competitive Accretion (CA)                & 4.26   & 3.26    & 7.05 & 4.73  
\enddata
\tablenotetext{a} {$\tau = 1$ Myr; $\avg{t_{\rm II}} = 2$ Myr; $\avg{\tf} \simeq \avg{\tf}_{\rm obs}/2.3= 0.24$ Myr.}
\tablenotetext{b} {$n=1$}
\tablenotetext{c} {$\rmd=3.6$}
\tablenotetext{d} {$\rmdca=3.2$}
\end{deluxetable*}

The mean luminosities in the accelerating cases are up to 30\% lower than in the fiducial
non-accelerating cases for a fixed value of $\avg{\tf}$
because the accelerating cases have more low-mass protostars.
However, for a given observed value of
$\avg{\tf}_{\rm obs}$, the mean luminosities are raised since their formation time is a factor $\simeq 2.3$ smaller,
as shown in equation (\ref{tfacc}). It is this adjustment that accounts
for the good agreement of the accelerating models with the observations.

Allowing the accretion rates to taper off during the accretion period
has a varying effect on the mean luminosities. In the IS case, the
accretion rate is no longer independent of mass,
increasing the mean luminosity.
In comparison, the other models accrete at higher rates at early times, lowering the mean luminosity. Adding
tapering also differentiates the models slightly at $m_u$. 
However, when the curves are re-normalized using the observational
completeness limit, the differences are minimized and in the TC and CA cases the mean
increases.
Among the tapered, non-accelerating accretion models, only the IS model falls
within error under the constraint that the
mean star formation time is 0.44 Myr.

\subsection{Median Luminosity}

While the mean is sensitive to the maximum
luminosity, which observationally may be prone to both over-estimation and statistical
fluctuations, the median serves as a better proxy for the peak of the
distribution. In this case, systematic errors in the spectra fitting and extinction corrections are
likely to dominate the error in the observed value.
Table \ref{tablemed} gives the medians for each of the cases for
$\mup=3\msun$
and $\tfobs=0.44$~Myr, 
where the observed median is  $1.5^{+0.7}_{-0.4}
~\lsun$. 
For the fiducial case, only the 2CTC and TC models are within
error. When tapered accretion is included, all but the IS model agree.
For an acclerating star formation rate, the untapered 2CTC, 2CTC, TC
and tapered CA and TC models are within error.

\begin{deluxetable*}{lllll}
\tablewidth{0pt}
\tablecolumns{5}
\tablecaption{Median Luminosity $(L_{\rm med}(L_\odot)$ with $\tfobs=0.44$ Myr; $L_{\rm med,\,obs}=1.5^{+0.7}_{-0.4}~L_\odot$.
\label{tablemed}}
\tablehead{ 
\multicolumn{1}{c}{} & \multicolumn{2}{c}{Non-Accelerating} & \multicolumn{2}{c}{Accelerating\tablenotemark{a}}\\ 
\colhead{Model} &\colhead{Untapered} & \colhead{Tapered\tablenotemark{b}}&\colhead{Untapered }& \colhead{Tapered\tablenotemark{b}}}
\startdata
Isothermal Sphere (IS)                    & 3.12  & 3.19  & 3.86  & 3.22 \\
2C Turbulent Core (2CTC)\tablenotemark{c} & 1.76  & 1.85  & 1.41  & 2.44  \\
Turbulent Core (TC)                       & 1.44  & 1.52  & 1.91  & 2.22 \\
2C Competitive Accretion (2CCA)           & 0.95  & 1.75  & 1.52  & 2.76 \\
Competitive Accretion (CA)                & 0.88  & 1.25  & 1.31  & 1.69
\enddata
\tablenotetext{a} {$\tau = 1$ Myr}
\tablenotetext{b} {$n=1$}
\tablenotetext{c} {$\rmd=3.6$.}
\end{deluxetable*}

\subsection{The Star Formation Timescale, $\avg{\tf}$}

We see that in some cases the mean and median luminosities are quite different from
the observed values when we fix the star-formation time at $\tfobs=0.44$~Myr. However,
this time scale is uncertain by a factor 2. Henceforth,
rather than fixing the observed lifetime, we derive the average formation time, $\avg{\tf}$, by
adjusting the models such
that in all cases the average model luminosity agrees with the average observed
luminosity, $\avg{L} = \avg{L_{\rm obs}}$.
Figure \ref{tffig} shows
the star formation timescale derived from the observed mean luminosity 
versus the 
mean model luminosity
obtained above 
from the observational timescale reported by \citet{evans09}. The
timescales are also listed in Table \ref{tableres}, which includes the
dimensionless parameter values for each model. 
For non-accelerating models, the mean formation time is the same as
the value that would be inferred from observation by comparing the number
of protostars with the number of Class II sources. For accelerating models,
however, the formation time is significantly less,
$\avg{\tf}\simeq \tfobs/2.3$, for the parameters we have adopted; both
values are listed in the table.
For models with 
non-accelerating accretion, 
the figure
strongly suggests that the actual formation time, during which the
majority of a protostar's mass is accreted, is
\beq
\avg{\tf}=0.3\pm0.1~~~\mbox{Myr}.
\eeq
This is longer than the observed Class 0 lifetime
of 0.1 Myr (e.g., \citealt{enoch09}), which implies that significant
accretion continues into the Class I phase.

Alternatively, one could use the median rather than the mean to adjust
the models. This would result in an inferred distribution of $\avg{\tf}$
more evenly distributed above and below $\avg{\tf}_{\rm  obs}$. 
For example, the untapered, nonaccelerating CA case would require a time half the length
of the observed formation time, while the untapered, nonaccelerating
IS case would require a time twice as large. However, adjusting the mean gives better agreement between the
overall distributions (see Section 4.5).

The vertical error bars on $\avg{\tf}_{\rm obs}$
arise from the uncertainty in the adopted disk lifetime: 2$\pm 1$ Myr.  
Another, smaller source of error arises from the possible misclassification of sources.
In comparing the observed formation time with that predicted by the models summarized
in Table 3, further uncertainty is introduced by the uncertainty in the mean luminosity
of the sample, which has been used to normalize the models.
Models without acceleration have an average formation time of about 0.3 Myr, which would
require disk lifetimes close to 1 Myr, smaller than most of the results cited by \citet{evans09}.
Models with an acceleration time $\tau=1$~Myr have an average observed formation
time of 0.6~Myr, corresponding to
a mean disk lifetime of about 3~Myr, at the high end of the observationally inferred
values. Improvements in the accuracy of the measured protostellar lifetimes,
acceleration times and luminosities will enable more stringent tests of the models.

\subsection{The Protostellar Luminosity Function (PLF)}

In this section we present the PLF for each model and define four 
dimensionless parameters to characterize the PLF shape. 
In Figures \ref{plfshapes} and \ref{plfshapes2}, we plot the predicted PLFs and overlay the PLF of
the observed distribution of protostars. 
We exclude sources with $L < 0.05 ~\lsun$ (comparable to the limit of the observations
after correction for extinction) 
from the model distributions and then renormalize the PLF to unity.

In Table \ref{tableres} we give the standard deviation of log luminosity, the ratio
of the median luminosity to the mean, the ratio of the maximum
luminosity to the mean, and the fraction of very low luminosity
objects (VELLOs; see Section 4.5.4) for each model, re-normalized
such that the means are equal to the observed mean luminosity. 

\begin{deluxetable*}{llcccccc}
\tablewidth{0pt}
\tablecolumns{8}
\tablecaption{Model Results for $\avg{L}=\avg{L_{\rm obs}}$
\label{tableres}}
\tablehead{ Model &  & $\sigma({\rm log}~L)$ & $\sigma_{\rm eff}({\rm log}~L)$\tablenotemark{a} &
  $L_{\rm med}/\avg{L}$ & $L_{\rm
  max}({\rm Class}\, 0)/\avg{L}$ & $f_{\rm VELLO}$\tablenotemark{b} & $\avg{\tf},\tfobs$(Myr)\tablenotemark{c} }
\startdata
Observed &  &$0.7^{+0.2}_{-0.1}$ & $0.7^{+0.2}_{-0.1}$& $0.3^{+0.2}_{-0.1}$
  & $10^{+4}_{-3}$ & $0.2 \pm 0.1$ & $0.44\pm0.22$ \\
\hline
\multirow{6}{*}{Isothermal Sphere (IS)}
& Non-Accelerating & &  & & & &\\
&~~~~~~Untapered                       & 0.44& 0.55 &0.96 & 5.81 &0.08 & 0.62\\
&~~~~~~Tapered\tablenotemark{d}        &0.45 & 0.56 & 0.83  &2.97 &0.08 & 0.72\\ 
&Accelerating\tablenotemark{e}& & & & & &\\
&~~~~~~Untapered                       & 0.43 &0.54 &0.86 & 2.90 &0.07
& 0.85, 0.37\\
&~~~~~~Tapered                         & 0.39 &0.51 &0.71 &3.67 &0.07
& 0.85, 0.37\\ \hline
\multirow{6}{*}{} 
&Non-Accelerating & & & & & &\\
&~~~~~~Untapered                       &0.55 &0.64 &0.60 &11.80 &0.09 & 0.56\\
Two-Component&~~~~~~Tapered            & 0.41 &0.53 &0.64 &5.23 & 0.08 &
0.60\\ 
Turbulent Core (2CTC)\tablenotemark{f} & Accelerating & & & & & &\\
&~~~~~~Untapered                       & 0.56 &0.65 &0.26 & 10.57&0.15
& 0.99, 0.43\\
&~~~~~~Tapered                         & 0.42 &0.53 &0.45 &6.74 &0.07
& 1.03, 0.45\\ \hline
\multirow{6}{*}{Turbulent Core (TC)}
&Non-Accelerating & & & & & &\\
&~~~~~~Untapered                      & 0.77 &0.84 &0.42 &9.46 & 0.18 & 0.69\\
&~~~~~~Tapered                        & 0.69 &0.76 &0.51 &4.91 &0.21 & 0.58\\ 
&Accelerating & & & & & &\\
&~~~~~~Untapered                      & 0.76&0.83 &0.32 & 9.78 &0.21 &
1.08, 0.47\\
&~~~~~~Tapered                        & 0.71&0.78 &0.53 &6.06 &0.12 &
0.84, 0.37\\ \hline
\multirow{6}{*}{}
&Non-Accelerating & & & & & &\\
&~~~~~~Untapered                      & 0.53 &0.61 &0.30 &9.46 &0.09 &0.55\\
Two-Component Competitive &~~~~ Tapered    & 0.49&0.57 &0.66 &4.04 &0.06 & 0.51\\ 
Accretion (2CCA)\tablenotemark{g}
&Accelerating & & & & & &\\
&~~~~~~Untapered                      & 0.55 &0.63 &0.30 &12.19 &0.11 & 0.94,0.41\\
&~~~~~~Tapered                        & 0.48 & 0.57 &0.67 &4.02 &0.05 & 0.78,0.34\\ \hline
\multirow{6}{*}{Competitive Accretion (CA)}
&Non-Accelerating & & & & & &\\
&~~~~~~Untapered                     &0.81 & 0.94&0.21 &12.78 & 0.25& 0.81 \\
&~~~~~~Tapered                       & 0.75 &0.89 &0.44 & 5.99&0.22 & 0.62\\ 
&Accelerating & & & & & &\\
&~~~~~~Untapered                     & 0.80& 0.93&0.18 &13.41 & 0.27 &
1.33, 0.58\\
&~~~~~~Tapered                       & 0.80& 0.93&0.43 &6.79 &0.17 &
0.89, 0.39\\ 
\enddata
\tablenotetext{a} {Effective standard deviation, which assumes a factor of
two time-variability in the observed luminosities; 0.3 dex is added
in quadrature to the model values given for $\sigma({\rm log}~L)$.}
\tablenotetext{b} {See equation \ref{eq:vello}.}
\tablenotetext{c} {For accelerating models, both the actual mean star-formation time,
$\avg{\tf}$, and the
one that would be inferred by comparing the number of protostars with the number of Class II sources, $\tfobs$, are given. These times are equal for non-accelerating models.}
\tablenotetext{d} {$n=1$}
\tablenotetext{e} {$\tau = 1$ Myr}
\tablenotetext{f} {$\rmd=3.6$}
\tablenotetext{g} {$\rmdca=3.2$}
\end{deluxetable*}

\subsubsection{Standard Deviation of Log Luminosity, $\sigma({\rm
  log}L)$}\label{stddev}

We find that the IS model has the smallest $\sigma({\rm log}L)$ in
all cases and is too narrow with respect to the data.  
In contrast, the TC and CA models have larger values of
$\sigma({\rm log}L)$ and encompass the extent
of the data.
The 2CTC and 2CCA models present a promising compromise 
and their $\sigma({\rm log}L)$ are slightly
too low.
Allowing the accretion rate of the protostars to taper off  
decreases the standard deviation of the
distributions in all but the non-accelerating IS case. This exception
occurs because tapering introduces a spread in the
accretion rate. 

Allowing for time variations
of a factor of two, i.e., non-FU-Ori fluctuations in the accretion
rates, would also increase the standard deviation but have little effect on
$\avg{L}$ or 
$L_{\rm med}/\avg{L}$.
To take such low-level variability into
account, we add 0.3 dex in quadrature to the
measured standard deviations. As shown in Table \ref{tableres}, 
this permits agreement between observations and both the 2CTC
and 2CCA models. The TC and CA models remain within error,
while the IS models continue to be inconsistent.

In Figure \ref{stdvsmu}, we plot $\sigma({\rm log}L)$ as a function of the most massive forming star.
We plot the tabulated values for the case of $\mup = 3~\msun$ and the 
observed standard deviation 
($\sigma({\rm log}L)= 0.7^{+0.2}_{-0.1}$ dex) in inset plots. 
As illustrated by both Figures \ref{stdvsmu} and \ref{plfshapes}, the standard deviations of the four models 
are very distinct. The TC and CA
models grow closer together for larger $\mup$, but remain fairly well
separated at $\mup = 3~\msun$. 
The standard deviations are also
relatively insensitive to the upper stellar mass in the cluster.
Tapering reduces the width of the distribution for all $\mup$
for these models.

\subsubsection{Median to Mean Ratio}

The ratio of the median to the mean, $L_{\rm med}/\avg{L}$, 
virtually eliminates the
dependence of the results on $\avg{\tf}$, $f_{\rm epi}$, and $f_{\rm acc}$. 
Table \ref{tableres} gives the values of the ratios for each
of the accretion models, where the observed ratio is $0.3 \frac{+0.2}{-0.1}$.
In all cases, the IS model can be ruled
out.\footnote{\citet{dunham10} find that the IS model as
 characterized by constant accretion onto a disk
  may still be consistent provided that the accretion
 from the disk onto the star is completely episodic. Here, we rule
 out the IS model where accretion from the disk onto the star is
 smooth.} 
The non-accelerating 2CTC  and the tapered 2CCA models
are outside the observational error bars.

\subsubsection{Maximum Luminosity, $L_{\rm max}$}

To characterize the maximum luminosity of the distributions, we
define:
\beq
L_{\rm max} = \int_{L'}^{L_u} \Psi(L_{\rm acc})~d {\rm ln} L = 1/N_*,
\eeq
where $\Psi(L_{\rm acc})$ is the PLF derived using the accretion luminosity.
For parity 
between the observations and models, we compare with the ratio of $L_{\rm
max}$ to $\avg{L_{\rm obs}}$. 
The most
luminous observed source has a bolometric luminosity of $76~\lsun$, which is
likely an upper limit and may
be over estimated by a factor of $\sim$ two. Consequently, we adopt
$54^{+22}_{-16} ~\lsun$ in comparing with the models.  
Importantly, this source is a Class 0 object, which suggests
that the luminosity chiefly arises from accretion. Thus, we derive
$L_{\rm max}$ from $\Psi(L_{\rm acc})$, which assumes that the interior luminosity is
small compared to the accretion luminosity.
We note that the values of $L_{\rm max}$ in Table \ref{tableres} are based on the actual
model dispersion, $\sigma$, not the enhanced value $\sigma_{\rm eff}$ that allows for
factor of 2 variability; allowance for such variability would
increase $L_{\rm max}$ somewhat.

According to Table \ref{tableres},
the 
untapered CA models have 
the highest $L_{\rm max}/\avg{L_{\rm obs}}$, followed by the .
2CCA and 2CTC cases and the TC ones. 
The observed maximum luminosity is 
$\sim 2-3$ times higher than for the tapered PLFs, but it is within
error of the untapered 2CTC, 2CCA, TC and CA cases. 
Without a correction for the model
mean luminosities, the difference between
the observed and model $L_{\rm max}$ would be a factor of 5-6. 
It is unclear whether this discrepancy is an artifact of the larger error of the higher
luminosity measurements, variable accretion, or fluctuations in the
star formation rate.

The untapered cases, even though directly consistent with $L_{\rm max}/\avg{L}$, are discrepant with the observations in a
different way. For untapered accretion, in which the accretion rates 
increase monotonically, the maximum luminosities are
achieved only when $m$ reaches its final value, $\mf$. Consequently, in order for the untapered
cases to be consistent with observations either the
Class I phase must have higher observed luminosities than the Class 0
phase or the Class 0
phase must last much longer is currently inferred from
observations. The tapered IS, 2CTC, and TC models, though lower as shown
by Table \ref{tableres}, have peak luminosities that occur midway
through the formation time. In contrast, the untapered 2CCA and CA
models, which agree better with the observations, achieve those
luminosities only towards the end of accretion, and thus 
do not appear to be consistent with observation.

\begin{figure}
\epsscale{1.0}
\plotone{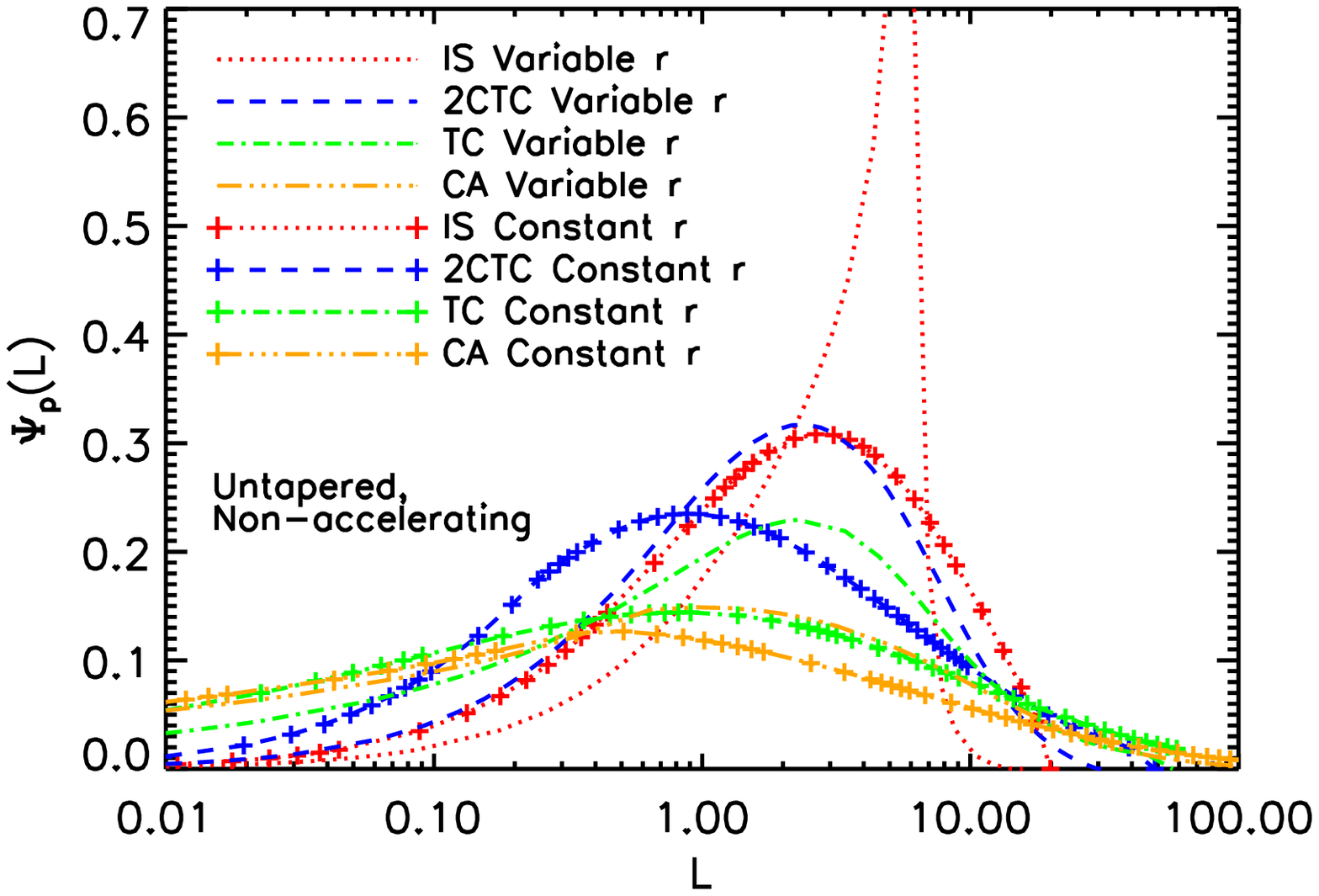}
\caption{ \label{plffigcomp} 
The PLF for $m_u = 3\msun$ for the fiducial cases, where $r$ is a
constant and where $r$ is a polynomial fit to $\bar
r(m)$. In each model, the curves are normalized to the same mean
luminosity, $\avg{L}$ (see \S \ref{meanl}).
}
\end{figure}

\subsubsection{Very Low-Luminosity Object Fraction, $f_{\rm VELLO}$}

There is observational evidence in support of a
significant population of low luminosity protostars. For
example, \citet{dunham08} found that $\sim$ 30\% of protostars with
$L \le 1.0~\lsun$
have luminosities below 0.1 $\lsun$, a sample commonly referred to as
very low-luminosity objects (VELLOs). In our sample, which covers
the same regions as \citet{dunham08} but  has been corrected for dust
extinction and more carefully pruned to eliminate non-protostars (e.g.,
background galaxies), only $\sim
20$\% of protostars can be considered VELLOs. Note that applying an exinction
correction to the data increases the median luminosity by 40\% so we define the
VELLO fraction as 
\beq
f_{\rm VELLO}=\frac{\caln_*(\lmin\leq L\leq 0.14 L_\odot)}{\caln(\lmin\leq L\leq 1.4 L_\odot)}.
\label{eq:vello}
\eeq
The observational sample contains one source with
an extinction-corrected luminosity of
$0.035~\lsun$, but it is likely incomplete at luminosities below
$0.05~\lsun$; we therefore set $\lmin=0.05 L_\odot$.
Without correcting for extinction, $\sim$ 20\% of the observed sources
would have $L \le 0.1 L_\odot$.

In Table \ref{tableres} we give $f_{\rm VELLO}$ for each of the models. 
The IS models, the tapered 2CTC, and the tapered 2CCA
models are inconsistent with the data, whereas the TC and CA models are consistent with the
observed values in all cases. 
The CA, and to a lesser extent, the TC models
also predict a substantial number of VELLOs below the
luminosity completeness limit. Future observations extending below
this limit are necessary to confirm or rule out these models.
It should be noted that in
all prescriptions, VELLOs are produced not by quiescent periods
preceeding episodic accretion events, but by 
the small accretion rates associated with protostars of very low mass. 
Current observations cannot
discriminate between low-luminosity sources in a quiescent
phase and those that are simply low-mass objects with low
accretion, although it is likely that at least some of the observed
protostars fall in the former category.
It is therefore possible that the low VELLO fractions of some of the models
may be consistent with the actual fraction of the subset of very low-luminosity protostars
that are not undergoing episodic accretion.

\subsubsection{Constant Radius PLF}
In the Appendix we derive the PLF for each model (except 2CCA) assuming only
accretion
luminosity and adopting a constant protostellar
radius, $r$.
Figure \ref{plffigcomp} shows these constant-radius curves in the
untapered, non-accelerating case together with the PLFs for which
$r$ is a polynomial fit to the sub-grid protostellar evolution
model (as in Figure \ref{plfshapes}).
Figure \ref{plffigcomp} illustrates that the curve width and maximum
luminosity are reduced when the radius is allowed to depend upon
mass. The constant radius curves have standard deviations that are
$\sim$ 50\% larger than when the radius is allowed to vary, a change
that, for example, makes the shapes of the constant radius IS curve and varying
radius 2CTC curves similar. 
This indicates that the stellar evolution model does
impact the PLF shape and the details of the comparison. In particular,
stellar evolutionary models with a narrower range of $r$ will have
broader luminosity distributions. Since our stellar evolution model depends upon
the instantaneous accretion rate and not just the accretion timescale, a star of
the same final mass will have a different amount of deuterium
remaining at $t=\tf$ for the different accretion histories. The models fall
in the order IS, 2CTC, 2CCA, TC, and CA from most
evolved to least evolved, based upon the amount of deuterium burned at
a given final mass. However, we note that the model PLF shapes are mainly
distinguished by their characteristic accretion models rather than 
differences in their stellar evolution.

\begin{figure}
\epsscale{0.95}
\plotone{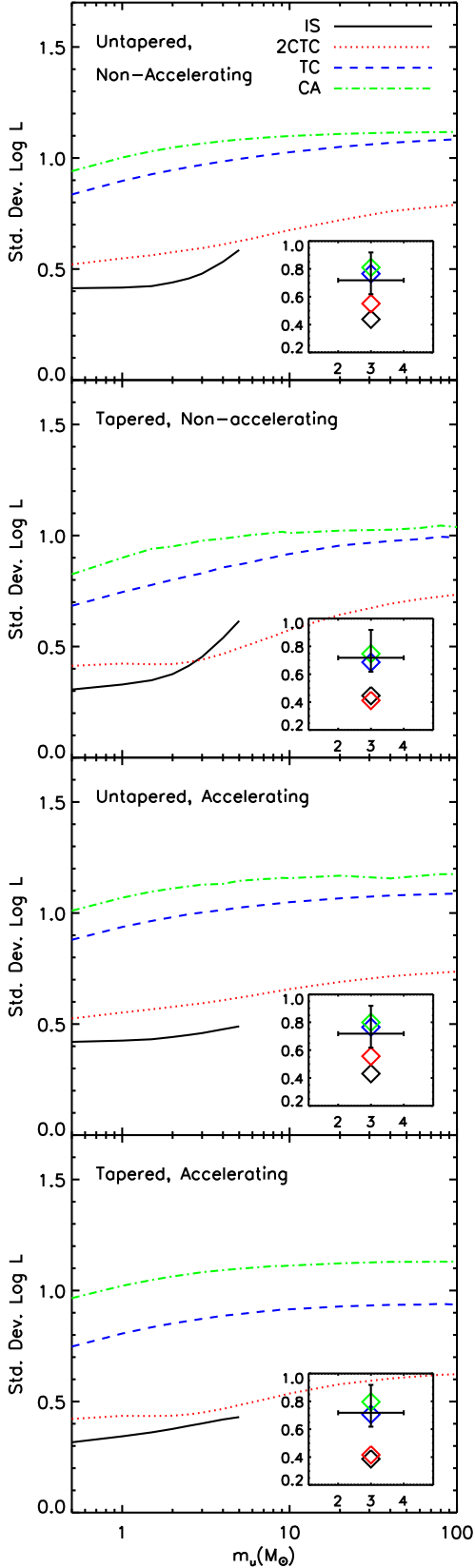}
\caption{ \label{stdvsmu} 
The standard deviation of the log of the protostellar luminosity as a function of
the upper protostellar mass, $\mup$, for the models with $\avg{\tf}=0.44$
Myr. For the tapered and accelerating cases $n=1$ and
$\tau=1$ Myr, respectively.
In the inset, the standard
deviation of the observed luminosities \citep{evans09} is plotted with error
bars to represent the uncertainty in the measurement and $\mup$. The
diamonds display the
values from Table 3, which assume $L_{\rm min} = 0.05~\lsun$ and
$\tfobs=0.44$.}
\end{figure}

\section{DISCUSSION} \label{s.dis}

A number of uncertainties enter into our comparisons. Foremost, the
observational data are difficult to obtain and corrections due to
obscuration along the line of sight contribute significant
error. There is also uncertainty in several of the parameters we have
adopted for the models and in the way these parameters are implemented. 

The parameter $f_{\rm acc}$ remains one of the most uncertain in our
estimation. 
It actually contains two very different effects: non-radiative loss of accretion
energy from the disk (e.g., \citealp{ostriker95}) and advection of accretion energy
into the stellar interior \citep{hartmann97}. The latter effect reduces the accretion luminosity by a factor $(1-\alpha)$, where $\alpha$ is the fraction of the accretion energy advected into the stellar interior. 
Most authors agree that this effect is small \citep{hartmann97,
baraffe09}, and it was not included in the work of
\citet{stahler88}. In a thorough study of the structure of
protostellar accretion shocks, \citet{commercon11} have shown that
$\alpha$ is indeed extremely small
provided the accretion flow is optically thin to optical radiation,
so we neglect it here. The accretion shock does heat the surface layers of the protostar, however,
and if one assumes that this heating is negligible because the accretion is localized onto
a small fraction of the protostellar surface, the resulting protostars can
be significantly more compact than when the surface of the protostar is heated by
the accretion shock \citep{hartmann97}. Such models
appear to be inconsistent with observation \citep{baraffe09, hosokawa11}.

The fraction of mass accreted during bursts, $f_{\rm epi}$, is also
uncertain. Most bursts are observed to occur in the Class I stage,
although Class II objects also experience FU-Ori type events
(e.g., \citealt{miller10}). There is some suggestion that bursts increase 
with age
and that Class 0 objects experience smoother accretion
\citep{vorobyov05, zhu09}. There is also evidence that isolated stars
experience more frequent outbursts than those in clusters
\citep{greene08}, a puzzle that highlights a deficit in our understanding
of protostellar accretion processes.

If protostars undergo periodic outbursts, then they must exist
in a quiescent, low-luminosity
stage between bursts. 
Indeed, some of these quiescent sources may appear as VELLOs. 
The median luminosity of the FU Ori sources cataloged by \citet{sandell01}
is $250 \,L_\odot$, so protostars that increase in luminosity by more than
8 magnitudes would have been in a VELLO state prior to outburst.
It is not known how many of the known FU Ori sources had pre-burst luminosities below
$0.14~\lsun$, but several of the best known ones did have pre-burst luminosities above this
limit \citep{hartmann96} and would not have been classified as VELLOs.
If the bursting sources were in a low-luminosity state between bursts,
they would populate the
lower luminosity region of the observed PLF and thus increase the 
standard deviation, lower the ratio of the median to the mean, and,
for sufficiently faint sources,
increase $f_{\rm VELLO}$
of the observed population relative to our models, which do not
explicitly include outbursts.
If the inter-burst accretion luminosity is negligible and the nuclear luminosity
is approximated by 
the ZAMS value, then non-accreting protostars with masses up to
about $0.7~\msun$ 
could potentially be classified as VELLOs.
The observed value of $f_{\rm VELLO}$ is thus an upper limit
on the number of sources in that luminosity range in our models.
As a result, models like the tapered
2CTC and 2CCA models, which fall below the observed $f_{\rm VELLO}$, are promising
candidate models, while the untapered TC and
CA accretion models may well over-predict $f_{\rm VELLO}$
and are not as promising.

The ratio of the median and mean luminosities 
can be significantly reduced by the degree of burstiness in the 
accretion rate of the individual protostars. 
The same effect can be achieved in models such as
the CA and TC models, which achieve a small ratio by virtue of a relatively
long phase of low-luminosity accretion.
However, as noted above, these models are somewhat artificial, the
former because protostars initially accrete some mass from a local
reservoir 
(which leads to the 2CCA model)
and the latter because turbulent motions do not exceed
thermal motions in such low-mass cores (which leads to the 2CTC model).
Nonetheless, this illustrates that it is possible to fit the observed
ratio without episodic accretion, and thus, that it is observationally difficult to
disentangle the influence of variability from an underlying mean
accretion trend shaping the luminosity distribution.

Although only a small number of sources have been observed to
undergo large, FU Ori outbursts, many protostellar objects have been
observed to undergo variability in luminosity over a factor of two on timescales of months
to years \citep{pech10, covey10}.
Our correction to the dispersion in log $L$ 
discussed in \S\ref{stddev} reflects this, but is very approximate.
In fact, we find that some low-level time-variability is needed in order for the
2CTC and 2CCA cases to be consistent with observations.

Another source of dispersion in the PLF could stem from variation in the initial conditions
for star formation. For example, temperatures in the low-mass star-forming regions we
study here are in the range $10-20$~K \citep{mckee10b}, which leads to
a $\sim$ factor 3 range of accretion luminosity in the IS model. This
corresponds to a dispersion of $\sim 0.15$~dex, which has a negligible
effect when added in quadrature to the dispersion of 0.3~dex for the
assumed temporal variability. When samples with a larger range of
initial conditions are considered, it is
possible that this additional source of dispersion would have to be included.

Our models do not directly take into account other modes of star formation
such as fragmentation within accretion disks
\citep{bate09a,stamatellos09, kratter10}.
Division of gas accreting onto a shared disk between close
companions is a complicated process, which could potentially alter the
the accretion dependence on $m$ and $\mf$ that we assume. 
However, the formulation of the models 
does not necessarily exclude disk fragmentation since we specify the
accretion prescription rather than the protostellar origin.
We also do not expect
disk fragmentation to be common in this observational sample, since heating due
to radiative feedback 
significantly stabilizes low-mass disks \citep{offner09, bate09}.
In the TC
and CA senarios, core and
filament fragmentation may produce wide binary companions
\citep{Offner10}, but accretion rates for this mode of
origin should follow the expected accretion trends.
We have not taken binarity into account in our analysis, but its effects are
small compared to the large differences among the different accretion models;
for example, the difference between Chabrier's (2005) IMF for individual stars
and that for stellar systems is only a factor of 1.25 in the peak mass.

In addition to the accretion history and to uncertainties in
parameters, inclination effects may also contribute to the standard deviation of the
observed PLF. 
For example, the extinction correction and therefore the 
bolometric luminosity of a protostar with a disk
that is observed edge-on will be underestimated. \citet{dunham10} found that adding
inclination effects to their models broadened the bolometric
luminosity distribution by $< 20$\%, so that orientation effects alone were not able to
fit the data. Moreover, including inclination effects had only a small
effect on the high luminosities, at most increasing the maximum
by 25\%. Consequently, we expect geometry to have a
minor effect on the shape of the PLF.

Adopting an accelerating star formation rate is one possible
resolution of the apparent discrepancy between 
the observed star formation timescale of 0.44~Myr,
which is based on the assumption of a constant star-formation rate \citep{evans09},
and the mean luminosity, which suggests a star-formation timescale of about 0.3~Myr
(Sec. 4.4).
We have adopted $\tau=1$~Myr as the acceleration time,
which is the inferred acceleration time for Ophiuchus and is likely to be 
a lower bound on the acceleration time for the other regions
\citep{palla00, mckee10}. 
(More recent Ophiuchus data suggest that the region has a deficit of
Class 0 objects and therefore a 
{\it decelerating} rate of star formation \citep{evans09}.)
Insofar as 1 Myr is a lower bound on the acceleration time, it gives 
the maximum difference
between $\avg{\tf}$ and $\avg{\tf}_{\rm obs}$, and thus
the maximum effect of acceleration on protostellar luminosities. 
IC348, a sub-region of Perseus, has an inferred
accleration time of 2 Myr \citep{palla00, mckee10}, but the accelerations for the
entire Perseus region and for Serpens are unknown. 
Given the small statistical sample size, the
appreciable uncertainties in the timescale estimates, and our
simplified acceleration model, these results should be interpreted as 
being consistent with, but not demonstrating that, acceleration
resolves the apparent discrepancy between
the observed star-formation time and the mean luminosity by
accelerating star formation.

How do the different accretion models compare with the data?
Independent of parameter uncertainties, 
the IS models appear to be inconsistent with observations. 
Of the five metrics we have adopted---$\sigma_{\rm eff}(\log L)$, $L_{\rm med}/\avg{L}$, 
$f_{\rm VELLO}$, $L_{\rm max}/\avg{L}$, and
$\tfobs$---only the last is within the
estimated errors. The ratio of the median to mean luminosity is more than 2 standard
deviations from the observed value. (However, the error estimates here are quite uncertain, and future data should substantially improve them.)
Because protostars accreting according to the TC and CA prescriptions 
spend significant time at low
masses, where IS-like accretion is likely to occur, 
we believe that the 2CTC and 2CCA models are more realistic physically,
even though in some cases the TC and CA models agree quite well with the data.
For example, in the CA case low-mass stars spend $1/3$ of
their accretion time achieving a mass of 0.1 $\msun$. It is unlikely that
protostars spend this much time at accretion rates significantly less
than the IS value. The 2CTC and 2CCA 
models are thus likely to be a better representation
of the TC and CA models, since the constant accretion
component dominates the initial protostellar
accretion rate.

Likewise, with the exception of the CA models, which are assumed to
have accretion terminated over a short time by protostellar feedback,
tapered models are likely to be more realistic since accretion from
a core declines gradually after the expansion wave reaches the surface
of the core (e.g., \citealp{mclaughlin97}).
Furthermore, untapered models achieve their maximum luminosity at the end of
the protostellar stage, whereas observations show that Class I sources are not noticeably
more luminous than Class 0 ones; in addition, the accretion rate for the Class 0
sources is believed to be larger than for the Class I sources, which is consistent
with tapering.
In our comparison we find that
in all cases the tapered models tend to underestimate
the maximum luminosity and
the standard deviation of the distribution. This may suggest that our
parametrization of tapering or our specific choice of $n=1$ is not
correct rather than ruling out tapered protostellar accretion rates.
It must also be borne in mind that we have not
included the effect of variability on
$L_{\rm max}$, so the tabulated values are lower bounds on the true
values.

As a check on our tapering model, we derive tapered PLFs for the IS
and TC cases assuming an exponential functional form for the accretion rate:
\begin{equation}
\dot m = {\mdo(m, \mf)} \exp(-2t/\tf)~~~~(t\leq \tf).
\end{equation}
For the IS case, this is similar to the model of
\citet{myers98}, except that in their model 
$m$ approaches $\mf$ as $t \rightarrow
\infty$. We modify their model so that $m=\mf$ at $t=\tf$,
at which point
the accretion rate has declined 
exponentially by a factor of $e^{-2}$.
For the IS case, we find that $\sigma({\rm Log}~L)$, $\avg{L}$,
$L_{\rm med}/\avg{L}$ and $L_{\rm max}/\avg{L}$ change by only a few
percent relative to the linearly tapered case. 
The parameter $f_{\rm vello}$ decreases by 50\%, mainly because the
exponentially tapered accretion rate does not go to $\sim0$ as the
linearly tapered case does. Thus, adopting a different tapering model
does not alter the fundamental disagreement of the IS case with the
observations. For the exponentially tapered TC case, $\sigma({\rm
Log}~L)$ and $L_{\rm max}/\avg{L}$ change by a few percent, while $\avg{\tf}$ and
$L_{\rm med}/\avg{L}$ increase by $\sim 20$ and $\sim 25$\%, respectively, and $f_{\rm vello}$
decreases by 40\%. Although these changes are more significant, the
dimensionless metrics continue to agree with the observations as
before,  except that $L_{\rm med}/\avg{L}$ becomes
slightly too high. While it may be possible to formulate accretion
tapering that improves the agreement with the observations, it seems
unlikely that a different functional form would modify our conclusion
that constant accretion time models agree better with the data than constant
accretion rate models.

The comparison with observations could be strengthened by deriving
PLFs for the Class 0 and Class I populations separately. 
In theory, the division between the
two classes occurs
when the protostellar mass exceeds the envelope mass
\citep{andre00,crapsi08}. 
For the IS and TC models, it is straightforward to
define Class 0 and Class I PLFs as the distributions of luminosities for which $m < \frac 12
\mf$ and 
$\frac 12 \mf \leq m \leq \mf$, 
respectively. Although such distributions can also be constructed for the CA
case using our approximate CA model, one characteristic of competitive accretion is the lack of correspondance between
envelope mass and final stellar mass \citep{smith09}.
Even using this simple definition, comparing the Class PLFs quantitatively
with the observations is challenging. 
Since the protostellar mass is not measurable during the embedded phase
and since inclination effects can significantly confuse the classification,
the two populations are difficult to distinguish observationally. 
Because the Class I lifetime is about
three times the Class 0 lifetime \citep{evans09},  models that
do not have a higher rate of accretion as Class 0 sources
could be excluded. If the fraction of a protostellar lifetime spent as a 
Class 0 source were to be less than that found by \citet{evans09}, 
the constraint on accretion and early luminosities would become more stringent.
Qualitatively, we expect all the untapered models
to fail such a test, since accretion, and hence luminosity, in such models rises
monotonically with age.

\begin{figure}
\epsscale{1.2}
\plotone{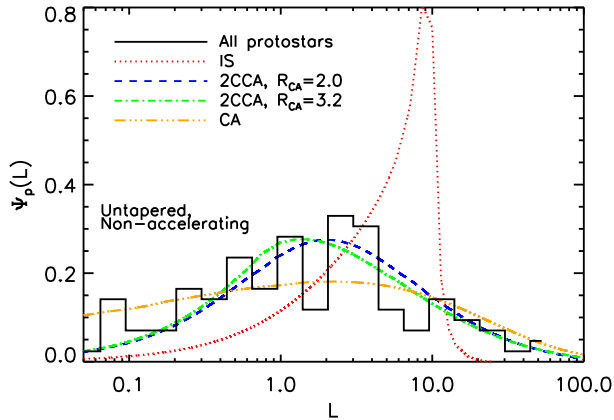}
\caption{ \label{plf2cca} 
The PLF for the untapered, non-accelerating star
formation 2CCA model with $\rmdca=2.0, 3.6$ and the corresponding 
IS, CA, and observational PLFs for comparison.}
\end{figure}

\section{CONCLUSIONS}

We have analyzed the Protostellar Luminosity Function (PLF), which is
the present-day luminosity distribution of a cluster of protostars,
for several different theories of star formation and compared the results
with observation. 
In our derivation we have assumed that the protostellar masses evolve
smoothly onto 
a truncated \citet{chabrier05} IMF, 
that the accretion rates are a
continuous function of the instantaneous and final protostellar
masses, and that the star formation rate is either constant or
accelerating in time.
We also assume that over most of the formation time, 
the accretion rate onto the protostar tracks the accretion rate from
the ambient molecular core onto the protostar-disk system. Episodic
accretion, such as that occurring in FU Ori outbursts, violates this 
assumption, but the available observational data indicate that this
is not a dominant effect.

The PLF depends explicitly upon the mean formation time of the
stars, $\avg{\tf}$, and the maximum stellar mass produced, $\mup$.  
For the low-mass star-forming regions we have compared with, $\mup$
appears to be about $3 M_\odot$.
We consider 
three main accretion prescriptions corresponding to standard
models of star formation: Isothermal Sphere (IS, constant accretion rate),
Turbulent Core (TC), and Competitive Accretion (CA, constant accretion time).
We note that prior to the development of either the CA or TC models,
\citet{kenyon90} considered both constant accretion rate and constant
accretion time in the paper that introduced the luminosity problem.
We also consider two hybrid models: the
Two-Component Turbulent Core model (2CTC, a compromise between the IS
and TC models) and the Two-Component Competitive
Accretion model (2CCA, a combination of the IS and CA accretion
prescriptions; this model was not considered in Paper I).
We explore two variations on these models: a case in which 
the accretion rate smoothly tapers to zero
and a case in which the star formation rate accelerates with a characteristic
timescale, $\tau$.

The CA model 
used here is an approximate analytic representation of the 
competitive accretion model developed by
\citet{bonnell97}, which begins with protostellar
seeds that are produced by a process similar to that in the IS case. As a result,
we believe the 2CCA model is a better approximation to their work
than the CA model. 
The TC model was specifically formulated for high-mass stars, which we do
not focus on here. For the case of low-mass stars, \citet{mckee03}
proposed the inclusion of an IS stage, which suggests that the 2CTC
model is the best representation of their model in this context. We
note that this model has some similarities to the TNT model of
\citet{fuller93}.

We compare our models to protostellar luminosities observed in local
low-mass star forming regions (Evans et al.~2009, Enoch et
al.~2009). 
The classical luminosity problem is that observed protostars appear to
have luminosities significantly lower than expected theoretically \citep{kenyon90}.
The extinction-corrected sample that we adopt from \citet{evans09}
has a mean luminosity of $5.3^{+2.6}_{-1.9}~\Lsun$, which is more than a factor of two larger
than the earlier \citet{enoch09} results that did not account for
extinction. This alone signficantly ameliorates the luminosity
problem. In comparing the models and the data, we 
first used the mean luminosity and the median luminosity. 
In all permutations of the parameters, the models require that the average
protostellar lifetime be $0.3\pm0.1$ Myr,
somewhat less than $> 0.4$ Myr
measured by \citet{evans09}. 
That is, the model luminosities are too {\it low} if the mean lifetime inferred
by \citet{evans09} is assumed. Thus, 
with a star formation time of $\sim 0.3$~Myr,
allowance for non-radiative energy loss
in winds ($\facc\simeq 0.25$) and a modest amount of episodic
accretion ($f_{\rm epi}\simeq 0.25
$) is sufficient to lower the mean protostellar luminosity
so that there is no longer a
``luminosity problem,'' in low-mass star formation.
We note that this resolution of the luminosity problem is quite consistent
with the suggestions of \cite{kenyon90}, who first pointed out the existence of the problem:
among the solutions they proposed were that the formation time was longer than
the $(1-2)\times 10^5$~yr indicated by their data and that some of the accretion was episodic.
With a star-formation time of 0.3~Myr, the mean accretion rate for a star of
mass $0.5 M_\odot$ (the mean mass of the \citealp{chabrier05} IMF) is 
$\dot m\simeq 2\times 10^{-6} M_\odot$~yr\e.

The discrepancy between our estimate of the mean star formation time
and that of \citet{evans09},
while not large,
could be due to a number of factors: the disk lifetime 
could be shorter than they assumed, 
the number of Class I sources could less than they assumed, 
$f_{\rm epi}$ or $\facc$ could be larger than we assumed,
and/or the star formation rate could be accelerating
as suggested by 
Follow-up studies of 
 dense gas tracers, such as HCO$^+$, have found that as
many as half of previously identified Class I objects are not actually embedded
\citet{kempen09, heiderman10}. Although \citet{evans09} exclude sources
embedded in less than 0.1 $\msun$ of gas mass when deriving the
lifetime, some sources may nonetheless be misclassified 
older, heavily obscured objects.
Also, while our value of $f_{\rm epi}$ is calculated using all the known
bursting sources, deeply embedded FU-Ori type sources may be 
missing from the sample.
By increasing the ratio of protostars to Class II sources, accelerating star formation
produces a longer observationally inferred star formation time than the actual star
formation time of the sample. Consequently, the accelerating models
have better consistency between the observed star formation time and
mean and median luminosities. Although we do expect the star formation
rate to vary with time, 
\citet{palla00} found
evidence for an
acceleration time as short as our adopted value, $\tau=1$~Myr, 
for only one the observed regions in the
protostellar sample (Ophiuchus), and this evidence is contradicted by
an apparent deficit of Class O sources in the region.

We then compared four dimensionless quantities that characterize the shape of the PLF:
(1) the standard deviation of
log $L$ (including an allowance for source variability at the factor of two level); 
(2) the ratio of the median to the mean luminosity; (3) the ratio
of the maximum to the mean luminosity; and (4) the fraction of
very low-luminosity objects, $f_{\rm VELLO}$, defined as the ratio of the number
of sources with extinction-corrected luminosities between $0.05 L_\odot$
and $0.14 L_\odot$ to the number between $0.05 L_\odot$
and $1.4 L_\odot$.
The first three of these are the most strongly
discriminating  since they are 
independent of $f_{\rm epi}$, $\facc$ and
the mean protostellar lifetime, $\avg{\tf}$;
the fourth quantity, $f_{\rm VELLO}$, is only weakly dependent on these factors.
We also compared the value of the 
star-formation time required by the models to get the observed mean protostellar 
luminosity with the star-formation time of
$\tfobs=0.44\pm0.22$~Myr inferred by
\citet{evans09} from the ratio
of the number of protostars to the number of Class II sources, which were assumed to have
a 2 Myr lifetime.
Although protostars with different accretion histories
have slightly different stellar evolutionary states at the end of accretion, we find that differences
between the PLF shapes are driven by the different accretion histories, not the different
evolutionary states. 
We find that the IS model is a poor fit to the data in all cases,
mainly due to the strongly peaked nature of its PLF profile. The
model results for the four dimensionless parameters are outside the error bars
of the data regardless of whether the model is tapered or untapered, accelerating or
non-accelerating. 
The one parameter the IS model agrees with is the observed star-formation time, after renormalizing the accretion rate so that the mean luminosity agrees with observation.

The CA model is in best agreement with the data; only the observed
star-formation time, $\tfobs$, lies outside the error bars, and this
is only for the
non-accelerating, untapered case.
The CA model predicts a relatively large number of VELLOs below the observational
limit of $0.05 L_\odot$, which will provide a strong test of the model in the future.
However, as discussed above, the 2CCA model, which has an initial phase in 
which the star accretes more rapidly, is closer to the actual competitive accretion
model.  Furthermore, one of the assumptions of the competitive accretion model is that the gas
is cleared out relatively quickly toward the end of the accretion phase, so the
untapered version of the 2CCA model is closest to the actual competitive accretion model.
However, 
this model (as well as the untapered CA model) predicts 
that the maximum luminosity
is achieved at late times, not during the Class 0 stage.
In general, the length of the Class 0 lifetime and the similarity of
Class 0 and Class I luminosities suggests that models that do not
accrete a significant fraction of mass during the earliest times are
inconsistent. This includes all the untapered models, which acheive
their maximum accretion rate, and hence maximum luminosity, at the end
of the protostellar lifetime, in what would be the late Class I stage.
Otherwise, both the accelerating and non-accelerating untapered versions of the 2CCA model agree
well with the data, although the dispersion is very slightly below the observed value for
the non-accelerating case.
The tapered version of the 2CCA model does not compare well with the data.

The TC models are in good agreement with the data, with the exception
of the non-accelerating, tapered case, which has 
a maximum-to-mean luminosity ratio that is marginally too low. The observed star-formation time,
$\tfobs$, is marginally too high for the non-accelerating, untapered case. As for the CA
model, however, the 2CTC model, which is similar to the IS model at low masses, is
more realistic. The untapered version of this model agrees well with the data, except that
the median luminosity is somewhat high for the non-accelerating case. However, the
tapered version of the 2CTC model is the best representation of the model, since
the model is essentially a turbulent version of the IS model. The non-accelerating
version of this model marginally agrees with all the data
except for $L_{\rm max}/\avg{L}$;
the accelerating version underpredicts the number of VELLOs (although as remarked above, this may not be a problem if some
of the observed VELLOs are episodic sources in a quiescent phase).

We conclude that models that tend towards a
constant accretion time and thus produce a greater spread in
luminosities (like the CA and TC models), rather than models
that have a constant accretion rate (such as the IS model) are in
better agreement with the data on the PLF.
Ultimately, agreement between a model and the observed PLF is 
necessary but not sufficient.
Models must also reproduce a number of other observed features of
protostars and the regions from which they form, including
core properties (e.g. \citealt{offner08, kirk09}),
the approximate agreement between the luminosities of Class 0 and Class I
sources, and the existence of large disks around Class I sources. The CA model, which
exhibits good agreement with the PLF, does not do well with
any of these additional features; in particular, it does not produce well-defined
cores for individual protostars \citep{enoch08}. 
The IS and TC models naturally have cores, and the tapered models can give
comparable luminosities for the Class 0 and Class I stages. Non-magnetic IS
and TC models are predicted to have large disks, but the existence of such
disks in the presence of magnetic fields is a topic of active investigation
(e.g., \citealp{mellon09,ciardi10}).

In this paper, we have developed the PLF as a tool for confronting star formation
theories with observation. Ongoing observational efforts should permit a significant
improvement in the comparison between theory and observation by providing
larger samples, which would reduce statistical fluctuations, and more accurate 
extinction-corrected luminosities, which would reduce the current factor of 2 uncertainties.
The sample of protostars we have analyzed has an estimated maximum mass 
$\mup=3 M_\odot$; a larger sample would presumably include more massive
stars and enable a stronger test of the theories.
If the sample were sufficiently large, one would be able to directly determine
the role of large, FU Ori type outbursts on the growth of protostars. Study of
the very faint protostars, the VELLOs, should determine the relative proportion
of those that are in a quiescent state between outbursts and those that are
faint because they have very low mass (as we have assumed). A monitoring program would
permit one to characterize the variabliity of the protostars, which we have
simply taken as increasing the dispersion in the PLF by a factor of two.
Finally, more accurate measurements of the physical conditions in the
star-forming clumps would enable more accurate theoretical predictions of
the accretion histories of the protostars in the sample.

\acknowledgements{We thank Steven Stahler, Jean Francoise Lestrad,
 Mark Krumholz, Lee Hartmann, Scott Kenyon, Joan Najita, Phil Myers
 and an anonymous referee for helpful comments. We
thank Melissa Enoch, Neal Evans, and Mike Dunham for useful discussions and clarifications of the observational data. This research has
been supported by the NSF through grants 
AST-0908553 (CFM) and AST-0901055 (SSRO)}.

\appendix

\section{PLF Alternative Derivation}

The PLF may be obtained either by integrating $\Ppt(L,m)$ over $m_f$ or
over $m$, where in either case the resulting PLFs are identical. Above we exclusively
use the former formalism. For completeness, we give the latter PLF
definition here: 
\beqa
\Ppl&=&\int d\ln m \Ppt(L,m),
\label{eq:ppl1}\\
&=&\int_\mmin^\mmax d\ln m\,\frac{\ppt[m,\mflm]}{\displaystyle \left|\ppbyp{\ln L}{\ln \mf}\right|},
\label{eq:ppl2}
\eeqa
where the lower limit of integration is given by equation (\ref{eq:mfl}) with $m=\mlmf$.
Note that this formulation does not work when the luminosity is independent of the
final mass, as in the case of Isothermal Accretion.

\section{The PLF for Constant Radius}

Accurate evaluation of the PLF requires allowing for the dependence of the
protostellar radius on the mass and accretion rate. However, the
radius is almost always within a factor 2 of $r=2.5~\rsun$, and if we
take $r$ to be constant the analysis is simplified considerably.

Combining equations ({\ref{eq:lacc}) and (\ref{eq:mdt}),
we express the accretion luminosity as
\beq
\lacc=L(1)\;\frac{m^{1+j}\mf^{\jf-j}}{r}\left[1-\delta_{n1}\left(\frac{m}{\mf}\right)^{1-j}\right]^{1/2},
\eeq
where
\beqa
L(1)&=&\facc G\mdoo\left(\frac{M_\odot^2}{\rsun \mbox{yr}}\right),\\&=&31.3\facc
\left[\frac{\mdoo}{1\times 10^{-6} \;\msun\;\mbox{yr\e}}\right]~~~\lsun,
\eeqa
is the luminosity for $r=1~\rsun$ and $m=\mf=1~\msun$ in the untapered case.
If we let
\beq
\ell\equiv\frac{rL}{L(1)}
\eeq
(where $r$ is in units of $\rsun$), then the relation of $m$ and $\mf$ to $L$ is
\beq
\ell=m^{1+j}\mf^{\jf-j}\left[1-\delta_{n1}\left(\frac{m}{\mf}\right)^{1-j}\right]^{1/2}.
\label{eq:ell}
\eeq

\subsection{Isothermal Sphere Accretion}

In this case, we have $m=\ell$, so that equation (\ref{eq:pplis3}) gives
\beq
\Ppl=\psi_p(m=\ell)=\frac{\ell}{\avg{\mf}}\int_{{\max(m_\ell,\ell)}}^\mup d\ln\mf\;\psi(\mf).
\label{eq:ppls2}
\eeq

\subsection{Tapered Isothermal Sphere Accretion}

In this case, equation (\ref{eq:ell}) for the accretion luminosity gives
\beq
\ell^2=m^2-\frac{m^3}{\mf},
\label{eq:ellis}
\eeq
which can be solved for $\mfellm$,
\beq
\mfellm=\frac{m}{\displaystyle 1-\frac{\ell^2}{m^2}}.
\eeq
Note that $m>\ell$. 
The result for the PLF is then (eq. \ref{eq:ppl2}),
\beq
\Ppl=\int_{\mmin}^\mmax d\ln m\;\frac{\displaystyle 2}{(m/\ell)^2-1}\;\ppt\left[m,\mf=\frac{m}{1-(\ell/m)^2}\right].
\eeq
The minimum possible value of $m$ for a given luminosity
corresponds to $\mf=m_u$. As $m$ increases, $\mf$ decreases until it
reaches a minimum, 
\beq
\min(\mf)=\left(\frac{3^{3/2}}{2}\right)\ell,
\label{eq:minmf}
\eeq
and it then increases again; as a result, the maximum
possible value of $m$ for a given luminosity also corresponds to $\mf=m_u$.
The limits of integration of the PLF, $\mmin$ and $\mmax$, are therefore
given by the roots of equation (\ref{eq:ellis}) with $\mf=m_u$ and that satisfy
the condition $m>\ell$. 
In order for $\mmin$ and $\mmax$ to be less than $m_u$, it is necessary
that $\min(\mf)$ be less than $m_u$, which sets an upper limit on the
luminosity,
\beq
\ell_u=\left(\frac{2}{3^{3/2}}\right)m_u.
\eeq
One can show that $\ell_u$ is the maximum luminosity allowed by equation
(\ref{eq:ellis}) with $\mf=m_u$.
It is also necessary that $\mf$ exceed $\ml$, which becomes an issue when
$\ell<2\ml/3^{3/2}$, so that $\min(\mf)<\ml$. In this case, the range of
mass between the roots of equation (\ref{eq:ellis}) with $\mf=\ml$ that satisfy
$m>\ell$ is excluded. The resulting range of integration extends from the
smaller root of equation (\ref{eq:ellis}) with $\mf=m_u$ to the smaller
root of this equation with $\mf=\ml$, and then from the larger root
equation (\ref{eq:ellis}) with $\mf=\ml$ to the larger root of the same
equation with $mf=m_u$.

\subsection{Untapered Turbulent Core and Competitive Accretion}

Since for untapered Turbulent Core and Competitive Accretion, the luminosity equation
(\ref{eq:ell}) implies
\beq
\mf=\left(\frac{\ell}{m^{1+j}}\right)^{1/(\jf-j)},~~~~m=\left(\frac{\ell}{\mf^{\jf-j}}\right)^{1/(1+j)},
\label{eq:mftcca}
\eeq
it follows that
\beq
\Ppl=\frac{1}{|\jf-j|}\int_\mmin^\mmax d\ln m\; \ppt\left[m,\mf=(\ell/m^{1+j})^{1/(\jf-j)}\right].
\eeq
The minimum mass for a given luminosity corresponds to the solution of
equation (\ref{eq:mftcca}) with $\mf=m_u$,
\beq
\mmin=\left(\frac{\ell}{m_u^{\jf-j}}\right)^{1/(1+j)}.
\eeq
There are two conditions that set the maximum value of $m$: First,
the minimum value of $\mf$ is $m_\ell$, so that
\beq
\mmax\leq \left(\frac{\ell}{m_\ell^{\jf-j}}\right)^{1/(1+j)}.
\eeq
Second, the requirement that $m$ be no more than the final mass, $\mf$, implies
\beq
\mmax\leq \ell^{1/(1+\jf)}.
\eeq
The correct value of $\mmax$ is the lesser of these two values.
The upper limit on the luminosity occurs when $m=\mf=m_u$,
\beq
\ell_u=m_u^{1+\jf}.
\eeq
The condition $\ell\leq\ell_u$ ensures that $\mmax\leq m_u$.

\subsection{Tapered Competitive Accretion}

In this case, we have
\beq
\ell=m^{5/3}\mf^{1/3}\left[1-\left(\frac{m}{\mf}\right)^{1/3}\right]^{1/2},
\label{eq:elltca}
\eeq
which leads to
\beq
\mf=\frac{m}{8}\left[1+\left(1+\frac{4\ell^2}{m^4}\right)^{1/2}\right]^3\equiv \frac{m}{8}(1+s)^3.
\eeq
Evaluating the partial derivative $\partial \ln L/\partial \ln\mf$, we find that the
luminosity function is
\beq
\Psi_p(L)=\int_\mmin^\mmax d\ln m \,\frac{12\ell^2}{m^4s(1+s)}\,\ppt\left[m,
\mf=\frac{m}{8}(1+s)^3\right].
\eeq
Just as in the case of tapered Isothermal Sphere accretion, there is a minimum
value of $\mf$ as a function of $m$ for a given luminosity,
\beq
\min(\mf)=\left(\frac{11^{11/4}}{10^{5/2}}\right)\ell^{1/2}.
\eeq
This is less than $m_u$ provided the normalized luminosity is less than
\beq
\ell_u=\left(\frac{10^5}{11^{11/2}}\right)m_u^2.
\eeq
For $\ell<\ell_u$, the limits of integration, $\mmin$ and $\mmax$,
are the roots of equation (\ref{eq:elltca}) with $\mf=m_u$ and $0<m<m_u$.

\subsection{Tapered Turbulent Core Accretion}

In this case, equation (\ref{eq:ell}) becomes
\beq
\ell=m^{3/2}\mf^{1/4}\left[1-\left(\frac{m}{\mf}\right)^{1/2}\right]^{1/2},
\label{eq:ellttc}
\eeq
so that
\beq
\mf=m\left(1+\frac{\ell^2}{m^{7/2}}\right)^2.
\eeq
The luminosity function is then
\beq
\Psi_p(L)=\int_\mmin^\mmax d\ln m \left(\frac{4\ell^2}{m^{7/2}+\ell^2}\right)\psi_p\left[m,\mf=m\left(1+\frac{\ell^2}{m^{7/2}}\right)^2\right].
\eeq
The limits of integration are the solutions of equation (\ref{eq:ellttc}) with $\mf=m_u$
and $0<m<m_u$, just as in the case of tapered Competitive Accretion. The maximum
possible luminosity can be found directly from maximizing $\ell$ in equation
(\ref{eq:ellttc}), or, equivalently, by evaluating $\min(\mf)$ and requiring that it
be less than $m_u$:
\beq
\ell_u=\left(\frac{6^3}{7^{7/2}}\right) m_u^{7/4}.
\eeq

\subsection{Two-Component Turbulent Core Accretion}

For Two-Component Turbulent Core Accretion, the accretion rate is
\beq
\dot m=\mdis \left(1+\rmd^2 m\mf^{1/2}\right)^{1/2}.
\label{eq:mdtctc}
\eeq
In this case, we define the normalized luminosity as
\beq
\ell\equiv\frac{rL}{L_{\rm IS}(1)}=m\left(1+\rmd^2 m\mf^{1/2}\right)^{1/2},
\eeq
so that the final mass for a protostar of mass $m$ and normalized luminosity $\ell$ is
\beq
\mf=\frac{1}{\rmd^4 m^2}\left(\frac{\ell^2}{m^2}-1\right)^2.
\eeq
Evaluating $\partial \ln L/\partial\ln\mf$, we find that the PLF is
\beq
\Psi_p(L)=\int_\mmin^\mmax d\ln m\; \frac{4}{1-(m/\ell)^2}\;
\psi_p\left[m,\mf=\frac{1}{\rmd^4 m^2}\left(\frac{\ell^2}{m^2}-1\right)^2\right].
\eeq
The lower limit of integration, $\mmin$, is the solution of equation (\ref{eq:mdtctc})
with $\mf=m_u$. Unless the luminosity is very low, the upper limit of integration is
set by the condition $m=\mf$, so that $\mmax$ is the root of the equation
\beq
\ell^2=\mmax^2\left(1+\rmd^2\mmax^{3/2}\right).
\eeq
However, $\mf$ cannot be less than the minimum stellar mass $m_\ell$, so
that in general $\mmax$ is the solution of the equation
\beq
\ell^2=\mmax^2\left[1+\rmd^2\mmax \max(\mmax^{1/2}, m_\ell^{1/2})\right].
\eeq
The maximum possible luminosity occurs when $m=\mf=m_u$, so that
\beq
\ell_u=m_u^2\left(1+\rmd^2 m_u^{3/2}\right).
\eeq
For $\ell<\ell_u$, both $\mmin$ and $\mmax$ are less than $m_u$.

\subsection{Tapered Two-Component Turbulent Core Accretion}

In terms of
\beq
g\equiv\left(1+\rmd^2 m\mf^{1/2}\right)^{1/2},
\eeq
the equation for tapered accretion is
\beq
\dot m=\mdis g\left(1-\frac{t}{t_f}\right).
\eeq
Integrating this equation gives the relation for the age as a function of mass,
\beq
t-\frac{t^2}{2t_f}=\frac{2}{\mdis\rmd^4\mf^{1/2}}(g-1).
\eeq
This relation shows that the formation time with tapering, $t_f$, is twice
the value without, $t_f=2t_{f0}$ with
\beq
t_{f0}=\frac{2}{\mdis\rmd^4\mf^{1/2}}(g_f-1).
\eeq
The tapering factor is then
\beq
1-\frac{t}{t_f}=\left(\frac{g_f-g}{g_f-1}\right)^{1/2},
\eeq
so that the normalized luminosity is
\beq
\ell=mg\left(\frac{g_f-g}{g_f-1}\right)^{1/2}.
\label{eq:elltctc}
\eeq
The protostellar luminosity function is then given by equation (\ref{eq:pplis2}) with
$\mf$ determined numerically from equation (\ref{eq:elltctc}) and with
\beq
\ppbyp{\ln L}{\ln\mf}=\frac{g-1}{8g^2g_f(g_f-g)}\left[(2g_f^2(g+1)
-3g(g_f-g)\right].
\eeq
Based on comparison with the cases of the tapered IS and tapered TC cases,
the upper and lower limits of integration for the PLF are found from numerical
solution of equation (\ref{eq:elltctc}) with $\mf=m_u$.

\bibliography{OffMck2010bib}
\bibliographystyle{apj}

\end{document}